\documentclass[12pt]{article}

\usepackage[utf8]{inputenc}

\usepackage{natbib}

\usepackage{amsmath, amssymb, amsfonts, amsthm, mathrsfs, bm}
\usepackage{color, float, graphicx, caption, subcaption}
\usepackage{comment}

\usepackage{mathtools}%for paired delimiters

\usepackage{hyperref}

\usepackage{setspace}
\graphicspath{{InkscapePics/}}

\newcommand{\executeiffilenewer}[3]{%
 \ifnum\pdfstrcmp{\pdffilemoddate{#1}}%
 {\pdffilemoddate{#2}}>0%
 {\immediate\write18{#3}}\fi%
}

\renewcommand{\a}{\alpha}

\newcommand{\e}{\varepsilon}
\newcommand{\g}{\gamma}

\renewcommand{\l}{\lambda}

\newcommand{\p}{\rho}

\newcommand{\z}{\zeta}

\newcommand{\D}{\Delta}

\newcommand{\R}{\mathbb{R}}

\newcommand{\ol}{\overline}

\newcommand{\ul}{\underline}

\renewcommand{\Pr}{\mathbf{P}}

\newtheorem{lemma}{Lemma}

\newtheorem{theorem}{Theorem}
\newtheorem*{polar}{Polar Factorization Theorem}
\newtheorem{proposition}{Proposition}
\newtheorem*{definition}{Definition}

\newtheorem{remark}{Remark}
\newtheorem{assumption}{Assumption}

\DeclareMathOperator*{\argmax}{arg\,max}

\allowdisplaybreaks

\begin{document}

\begin{titlepage}
\centering{\Large \textbf{INTERVENTIONS AGAINST MACHINE-}}\\[.2cm]
{\Large \textbf{ASSISTED STATISTICAL DISCRIMINATION}}\\[1cm]
\centering{\Large \textbf{}}\\[1cm]
{\large JOHN Y. ZHU}\footnote{University of Kansas, johnzhuyiran@ku.edu. I thank Li Hao, Hyunseob Kim, Kris Nimark, and Jennifer Ifft for reading through the manuscript and providing detailed feedback. I also thank Xiaosheng Mu, Fedor Sandomirskiy, Nathan Yoder, and seminar and conference audiences at Kansas State University, the NSF/CEME Decentralization Conference on Mechanism Design with AI and Distributed Ledgers, Princeton, the Kansas Workshop in Economic Theory, the North American Summer Meeting of the Econometric Society, the Midwest Economic Theory Conference, the University of Queensland, the Designing for Redistribution Conference, Pittsburgh-Carnegie Mellon, the University of Western Ontario, and the University of British Columbia for helpful comments and discussions. I am grateful to Crystal Lenz for excellent editorial assistance. All remaining errors are my own.}\\[.4cm]
\today\\[2cm]

\begin{abstract}
\noindent{}I study statistical discrimination driven by \emph{verifiable beliefs}, such as those generated by machine learning, rather than by humans. When beliefs are verifiable, interventions against statistical discrimination can move beyond simple belief-free designs, like affirmative action and blinding, to more sophisticated belief-contingent ones. I analyze a belief-contingent intervention, common identity, and show that it can be more effective at combating statistical discrimination than popular alternatives -- particularly when the training dataset exhibits the kinds of statistical biases that often plague machine-assisted decision problems.
\end{abstract}%12, 11, 9, 9, 8, 13, 9, 10 = 81
\begin{flushleft}
\end{flushleft}
\end{titlepage}
\newpage
\setcounter{page}{1}

\section{Introduction}

In many settings, a decision maker (DM) must act on an individual without observing a relevant quality of that individual. For example, employers and colleges make offers without observing who has high ability. Banks lend without observing who will default. Judges grant bail without observing who will commit a violent crime, if released. When an individual's quality is unobserved, the DM must infer said quality from ``features" of the individual that are observed, such as test scores, income, or zip code.

Suppose the DM believes one group of individuals, say, $B$, has worse average quality than another group, $A$. Then, in comparing two individuals with identical features, one from each group, the DM will form a worse belief about the group $B$ individual's quality. A vicious cycle of \emph{statistical discrimination} can now emerge: The DM's worse beliefs about quality in group $B$ cause the DM to apply a tougher decision policy to the group, such as a higher test score cutoff for acceptance. Group $B$ individuals then have less incentive to invest in quality. This results in a worse distribution of quality throughout the group, which ultimately fulfills the DM's worse beliefs about the group as a whole. Consequently, group $B$ gets stuck in a worse equilibrium with the DM than group $A$.

To combat statistical discrimination, policymakers often mandate some form of equal treatment for both groups. For example, affirmative action requires that the DM accept from both groups at the same rate, while group-blinding bans the DM from conditioning acceptance on group membership. While the specific form of equal treatment may differ across interventions, the intent is the same -- to give both groups the same incentive to invest in quality. The idea is that equal incentives lead to equally qualified groups. When group $B$ becomes just as qualified as group $A$, the DM no longer desires to treat group $B$ any worse than group $A$. The intervention then ceases to be a binding constraint on the DM and can be lifted. In this way, equilibria between the two groups and an intervention-constrained DM coincide with scenarios in which both groups play the same equilibrium with an unconstrained DM.

Despite their best intentions, many popular equal treatment interventions fail to guarantee equal incentives. Consider affirmative action. When group $B$ is less qualified than group $A$, the need to equalize acceptance rates can actually cause the DM to go from being too tough to too soft on group $B$ \citep{coate1993will}. Either way, group $B$ has less incentive to invest in quality than group $A$.

Or, take group-blinding. If the only feature is the score on a test of quality, then group-blinding induces equal incentives, since the DM must apply the same test score cutoff for acceptance to both groups. But if, in addition to the test score, there is a second feature that depends only on group membership, then a group-blinding-constrained DM can still provide the two groups different incentives -- for example, by setting a higher test score cutoff for acceptance, whenever the second feature takes values more likely to be generated by group $B$.

Neither affirmative action nor group-blinding conditions on the DM's beliefs about quality, despite the crucial role those beliefs play in statistical discrimination. The reason many interventions do not condition on beliefs is because, historically, beliefs were unverifiable thoughts formed in a human's mind. However, the rise of big data has led many decision makers to rely on machine learning to compute their beliefs. Today, machine learning algorithms are used to assess ability in hiring and admissions, default risk in lending, and recidivism risk in pretrial bail decisions. Machine-learning-generated beliefs are verifiable. This makes possible belief-contingent interventions.

In this paper, I analyze an intervention, which I call \textbf{common identity}, that mandates a form of equal treatment hinging on the DM's beliefs about quality. Unlike affirmative action and group-blinding, common identity guarantees equal incentives. As a result, common-identity-constrained equilibria coincide with scenarios in which both groups play the same unconstrained equilibrium.

To fix ideas, suppose there are only two unobserved quality levels $\{$\textit{qualified}, \textit{unqualified}$\}$ and two decisions $\{$\textit{accept}, \textit{reject}$\}$. The DM only wants to accept qualified individuals, while all individuals want to be accepted. Let $X$ denote the set of features. Let $f_A : X \rightarrow [0, 1]$ summarize the DM's beliefs about quality in group $A$, where $f_A(x)$ is the probability the DM believes a group $A$ individual with features $x$ is qualified. Define $f_B$ similarly. Common identity picks an arbitrary reference group, say, $A$, and assigns all individuals with features $x$ the same ``summary score" $f_A(x)$, regardless of their group. The DM is then required to select a summary score cutoff and accept an individual if and only if their summary score exceeds the cutoff.

To gain intuition for the specific form of equal treatment mandated by common identity, imagine the DM has already accepted a group $A$ individual with some features $x$ and is about to act on a group $B$ individual with some different features $x'$. Complying with common identity requires the DM to probe their beliefs by asking:
\begin{quote}
Given the factual $(B, x')$ individual before me, how would I compare a counterfactual $(A, x')$ individual against the $(A, x)$ individual I accepted?
\end{quote}
If the DM thinks more highly of the counterfactual $(A, x')$ individual, then common identity requires that the factual $(B, x')$ individual be accepted.

Common identity induces equal incentives, even when the two groups are not equally qualified (unlike affirmative action) and even when some features depend only on group membership (unlike group-blinding). To see why, let an individual's features consist of two components, $x = (x_Q, x_G)$, where $x_Q$ depends only on quality, and $x_G$ depends only on group membership. When group $B$ is less qualified than group $A$, a common-identity-constrained DM selects a higher summary score cutoff than they would otherwise select if it only applied to group $A$. But once a cutoff is selected, the same rubric $f_A(\cdot)$ is applied to both groups. When group membership is held fixed, $x_G$, which depends only on group membership, is uninformative about quality. In particular, $f_A(x_Q, x_G) = f_A(x_Q)$. Since $x_Q$ depends only on quality and not group membership, both groups have the same incentive to invest in quality.

Common identity uses the beliefs about quality in a specific group. By contrast, let $f_{blind} : X \rightarrow [0, 1]$ denote the beliefs about quality that would be formed if one could only condition on an individual's features and not their group. A group-blinding-constrained DM is better off accepting all individuals with features $x$ if and only if $f_{blind}(x)$ exceeds some probability. Thus, group-blinding induces the same best-response from the DM as a modified version of common identity, where the summary score rubric is $f_{blind}(\cdot)$ instead of $f_A(\cdot)$. Given this connection, why does the intuition for common identity inducing equal incentives not apply to group-blinding? 

When there is statistical discrimination against, say, group $B$, group membership predicts quality. Because $f_{blind}$ cannot condition on group membership, some of group membership's predictive power is reassigned to features that correlate with group membership -- namely, $x_G$. This causes $f_{blind}$ to exhibit \emph{omitted variable bias}. For example, suppose $x_G \in \{\hat{A}, \hat{B}\}$, with group $A$ ($B$) more likely to generate $\hat{A}$ ($\hat{B}$). Then for all $x_Q$, we have $f_{blind}(x_Q, \hat{A}) > f_{blind}(x_Q, \hat{B})$, despite $f_A(x_Q, \hat{A}) = f_A(x_Q, \hat{B})$ and $f_B(x_Q, \hat{A}) = f_B(x_Q, \hat{B})$. If $x_Q$ is the score on a test of quality, then the set of test scores high enough to push $f_{blind}$ above the point where acceptance becomes optimal shrinks when $x_G$ changes from $\hat{A}$ to $\hat{B}$. This means a group-blinding-constrained DM optimally sets a higher test score cutoff for acceptance when $x_G = \hat{B}$, compared to when $x_G = \hat{A}$. Group $B$ individuals are then less likely to be accepted than group $A$ individuals, even after controlling for quality. Equal incentives are therefore not guaranteed.

Of course, if policymakers knew that $x_G$ depends only on group membership, they could also ban the DM from conditioning on $x_G$. The problem with such a blinding intervention is that the feature-generating process is often unknown, making it difficult to tell if a particular feature is part of $x_G$.

Instead, practical blinding interventions use a mix of social norms and statistical tests to determine when conditioning on a feature violates the spirit of group-blinding. Such ``contentious" features are then banned in addition to group membership (e.g., zip code when group membership is by race). Against this ad hoc regulatory backdrop, \citet{pope2011implementing}, \citet{ghili2019eliminating}, and \citet{yang2020equal} independently developed common identity, which is referred to as the ``proposed procedure" in \citet{pope2011implementing}, ``train-then-mask" in \citet{ghili2019eliminating}, and ``minorities-as-whites" in \citet{yang2020equal}. In a series of applications, they showed how decisions are more accurate -- fewer qualified (unqualified) individuals rejected (accepted) -- under common identity than under various interventions banning group membership and contentious features. In fact, accuracy under common identity is often almost as high as when only group membership is banned. At the same time, the acceptance rate gap suffered by the discriminated group is smaller under common identity than when only group membership is banned. I complement these studies by investigating the equilibrium impact of common identity on statistical discrimination.

Statistical discrimination as a vicious cycle was first articulated in \citet{myrdal1944american} and later modeled by \citet{arrow1973theory}. My paper builds on the canonical model of statistical discrimination by \citet{coate1993will}. A similar model was contemporaneously developed by \citet{foster1992economic}. More recent contributions to the literature have emphasized interactions between individuals \citep{moro2004general, onuchic2023signaling}, search frictions \citep{rosen1997equilibrium, mailath2000endogenous, jarosch2019statistical}, rational inattention \citep{fosgerau2023equilibrium, echenique2025rationally}, and learning traps \citep{komiyama2024statistical, li2025hiring, bardhi2025early}. See \citet{fang2011theories} and \citet{onuchic2025recent} for excellent surveys.

The type of decision problem common identity is designed for is known as binary classification in machine learning. In binary classification, there are two unobserved qualities, with the DM having a different preferred decision attached to each one. In many applications of binary classification, practitioners became concerned that the distribution of decisions across groups was imbalanced. The algorithmic fairness literature in computer science emerged to address these concerns.

Specifically, the literature formulated various \emph{group fairness} concepts that attempt to capture what it means for the distribution of decisions to be balanced across groups. Examples include equal probability of being accepted (referred to as statistical parity in machine learning), equal probability of being accepted conditional on being qualified (equality of opportunity), equal probabilities of being accepted conditional on being qualified and unqualified (equality of odds), and equal probability of being qualified conditional on being accepted (predictive parity). Various belief-contingent interventions were then designed to satisfy these group fairness concepts \citep{zemel2013learning, feldman2015certifying, hardt2016equality}. 

Social scientists have also contributed to the algorithmic fairness literature by highlighting the role of incentives. Scholars initially focused on how the DM's actions affect the feature distribution through their impact on the incentives of the individuals being acted upon \citep{hardt2016strategic, eliaz2019model, hu2019disparate, kleinberg2020classifiers, perdomo2020performative, frankel2022improving}. Their findings then led to work on the long-term implications of group fairness interventions \citep{hu2018short, mouzannar2019fair, d2020fairness, jung2020fair, liu2020disparate, puranik2022dynamic}. More recently, attention has turned to the incentives of the DM. \citet{fu2022fair} explore how group fairness interventions impact the DM's incentive to improve the accuracy of their machine learning algorithm. \citet{liang2024algorithm} and \citet{strack2024privacy} explore how to induce the DM to act in ways that satisfy various group fairness concepts through belief-contingent information design.

My paper both extends and departs from the algorithmic fairness literature. My analysis of the equilibrium impact of common identity factors in the incentives of both the DM and the individuals being acted upon. In this way, I take the logical next step from the articles that focus only on the incentives of one party. However, my goal differs from the group fairness goals of the algorithmic fairness literature because I aim to ensure both groups play the same unconstrained equilibrium, which is motivated by the theory of statistical discrimination.

\section{Model}\label{model}

There is a population of individuals with unit mass. Initially, each individual is assigned a group $g \in \{A, B\}$ and independently draws a cost $\in \R$. Let $\l_g \in (0, 1)$ be the fraction of the population assigned to group $g$. Everyone draws their cost from the same distribution, which has a strictly increasing and continuous cumulative distribution function (CDF) $H$. Next, each individual chooses their quality: Stay unqualified, $u$, for free or become qualified, $q$, at their drawn cost. The assumption that $H$ is strictly increasing over $\R$ implies $H(0) > 0$, which means some individuals draw negative costs and, therefore, gain utility from becoming qualified. While not crucial, the assumption ensures that, in any equilibrium, no group ever has only qualified or only unqualified individuals. The DM's equilibrium beliefs about quality are then never degenerate, which makes the analysis cleaner.

After quality is chosen, each individual generates a vector of features $x \in [0, 1]^N$, where $N$ is the number of feature types. For now, assume $x$ depends only on an individual's quality and not their group. Let $p_q(x)$ and $p_u(x)$ denote the full-support densities of $x$ conditional on being qualified and unqualified, respectively.

\begin{figure}[t]
  \centering
    \centering
    \scalebox{.66}{%
 \executeiffilenewer{InkscapePics/timeline.svg}{InkscapePics/timeline.pdf}%
 {inkscape -z -D --file=InkscapePics/timeline.svg %
 --export-pdf=InkscapePics/timeline.pdf --export-latex}%
 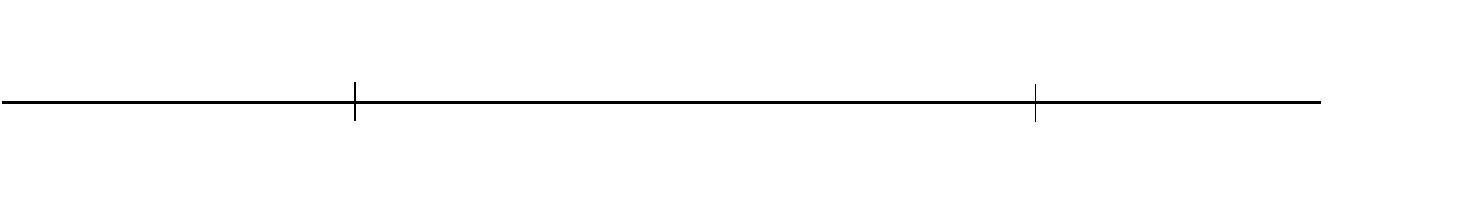%
}
    \caption{Timeline of the model.}
    \label{fig: timeline}
\end{figure}

There is a decision maker (DM) who observes each individual's group and features but does not observe quality. The DM begins by forming beliefs about quality in each group. The beliefs about quality in a group are summarized by a function $f : [0, 1]^N \rightarrow [0, 1]$, where $f(x)$ is the probability the DM believes a member with features $x$ is qualified.

After forming beliefs about quality, the DM applies a decision policy to each group. A decision policy is a map $d : [0, 1]^N \rightarrow \{1, 0\}$, where 1 means accept and 0 means reject. If an individual is accepted, they receive payoff 1, while the DM's payoff depends on the individual's quality, $v_q > 0$ or $-v_u < 0$. If an individual is rejected, both parties receive payoff 0. See Figure \ref{fig: timeline}.

Fix a group. Let $\mu$ be the density of the distribution of features in the group, and let $f$ be the DM's beliefs about quality in the group. Then the DM's utility from applying decision policy $d$ to the group is
\begin{align*}
V(d, \mu, f) := \int_{[0, 1]^N} d(x) \mu(x) \left[ f(x)v_q - \left(1- f(x)\right) v_u\right] dx.
\end{align*}

\begin{definition}
Given decision policy $d$, the incentive to become qualified is $Inc(d) := \int_{[0, 1]^N}d(x) \left( p_q(x) - p_u(x) \right) dx$, which is the difference in expected payoff between becoming qualified and staying unqualified facing $d$.
\end{definition}

Suppose decision policy $d$ is applied to a group. Then the best-response of a member is to become qualified if and only if their cost is less than or equal to $Inc(d)$. More generally, given any $c \in \R$, let $\ol{c}$ denote the profile of quality choices in a group, in which a member becomes qualified if and only if their cost is less than or equal to $c$. When a group's quality profile is $\ol{c}$, the fraction of qualified individuals in the group is $H(c)$, and the group's feature density is $\mu \vert H(c)$, where $\mu(x)\vert H(c) := H(c) p_q(x) + (1-H(c)) p_u(x)$.

\begin{definition}
An equilibrium between a group and an unconstrained DM -- or simply an unconstrained equilibrium -- consists of a quality profile $\ol{c^*}$, beliefs about quality $f^*$, and a decision policy $d^*$, satisfying the following conditions:
\begin{itemize}
  \item The quality profile is the best-response to the decision policy, $c^* = Inc(d^*)$.
  \item The beliefs about quality are rational given the fraction of qualified individuals,
  \begin{align*}
   f^* \equiv f\vert H(c^*), 		& \mbox{\ \ where\ \ } f(x)\vert H(c^*) := \frac{H(c^*) p_q(x)}{H(c^*) p_q(x) + (1-H(c^*)) p_u(x)}.
  \end{align*}
  \item The decision policy is a best-response given the feature density and the beliefs about quality, 
  \begin{align*}
  d^* \in \argmax_{\mbox{decision policies $d$}} V(d, \mu \vert H(c^*), f^*).
  \end{align*}
\end{itemize}
\end{definition}

Since the two groups are ex-ante identical, they have the same set of unconstrained equilibria. Let us characterize these equilibria and show they are Pareto-ranked. The analysis is adapted from \citet{coate1993will} Section I.

\begin{definition}
Given features $x \in [0, 1]^N$, the likelihood of $x$ is $\frac{p_q(x)}{p_u(x)}$. Given $l \in [0, \infty]$, the likelihood-cutoff decision policy $\ul{l}$ satisfies $\ul{l}(x) = 1$ if and only if $\frac{p_q(x)}{p_u(x)} \geq l$.
\end{definition}

Fix a group with quality profile $\ol{c}$, and let the DM's beliefs about quality in the group be rational. Then the DM is better off accepting a member with features $x$ if and only if $(f(x) \vert H(c)) v_q - (1 - f(x) \vert H(c)) v_u \geq 0$. This inequality simplifies to
\begin{align*}
\frac{p_q(x)}{p_u(x)} \geq L(c) := \left[ \frac{1 - H(c)}{H(c)}\right] \frac{v_u}{v_q}.
\end{align*}
So, the DM's best-response is the likelihood-cutoff decision policy $\ul{L(c)}$. Conversely, the group's best-response to a likelihood-cutoff decision policy $\ul{l}$ is the quality profile $\ol{C(l)}$, where $C(l) := Inc(\ul{l})$. This implies

\begin{lemma}
A quality profile $\ol{c^*}$ and a likelihood-cutoff decision policy $\ul{l^*}$ comprise an unconstrained equilibrium if and only if $c^* = C(l^*)$ and $l^* = L(c^*)$ -- that is, $(c^*, l^*)$ is an intersection of the graphs of $C$ and $L$.
\end{lemma}

In describing an unconstrained equilibrium, there is no need to specify the DM's rational beliefs about quality, which are determined by the quality profile.

Because $H$, the CDF of the cost distribution, is strictly increasing and continuous, with range $(0, 1)$, $L$ is strictly decreasing and continuous, with range $(0, \infty)$. Intuitively, the higher is $c$, the more qualified is the group, and the more lenient the DM wants to be, which translates to a lower likelihood cutoff for acceptance.

By definition, $C(l) = \int_{\left\{x \vert \frac{p_q(x)}{p_u(x)} \geq l\right\}} \left( p_q(x) - p_u(x) \right) dx$. When $l < 1$ ($> 1$), marginally increasing $l$ removes negative (positive) values of $p_q(x) - p_u(x)$ from the integral. Thus, $C$ is single-peaked, attaining its maximum at $l = 1$, and its minimum, 0, at $l = 0, \infty$. Intuitively, accepting everyone or no one provides zero incentive to become qualified. The incentive to become qualified is maximized when it is difficult, but not too difficult, to get accepted. To ensure $C$ is positive over $(0, \infty)$ and continuous, it suffices to assume the likelihood $\frac{p_q(x)}{p_u(x)}$, viewed as a function over $[0, 1]^N$, has range $[0, \infty]$ and does not map any set with positive Lebesgue measure to a single value.

\begin{figure}[t]
  \centering
    \centering
    \scalebox{.8}{%
 \executeiffilenewer{InkscapePics/unconstrained_eq.svg}{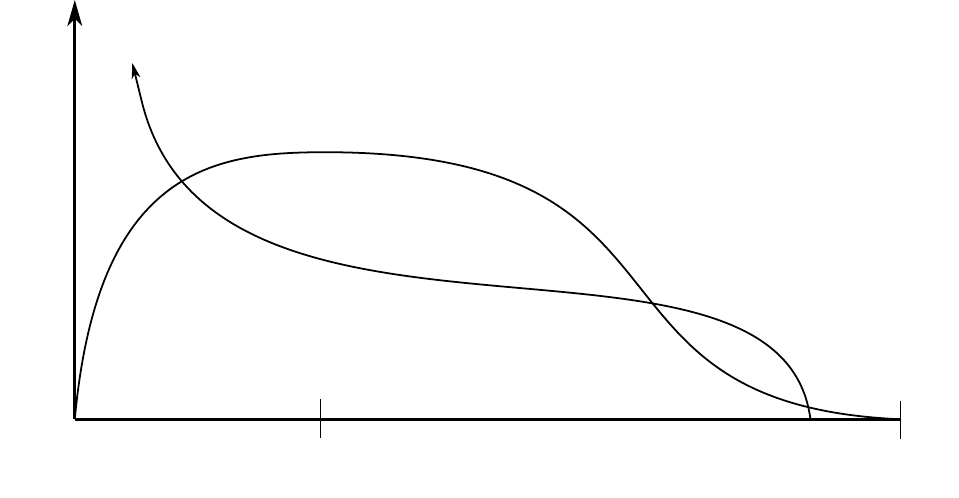}%
 {inkscape -z -D --file=InkscapePics/unconstrained_eq.svg %
 --export-pdf=InkscapePics/unconstrained_eq.pdf --export-latex}%
 %% Creator: Inkscape 1.0.2 (e86c8708, 2021-01-15), www.inkscape.org
%% Accompanies image file 'unconstrained_eq.pdf' (pdf, eps, ps)
\begingroup%
  \makeatletter%
  \providecommand\color[2][]{%
    \errmessage{(Inkscape) Color is used for the text in Inkscape, but the package 'color.sty' is not loaded}%
    \renewcommand\color[2][]{}%
  }%
  \providecommand\transparent[1]{%
    \errmessage{(Inkscape) Transparency is used (non-zero) for the text in Inkscape, but the package 'transparent.sty' is not loaded}%
    \renewcommand\transparent[1]{}%
  }%
  \providecommand\rotatebox[2]{#2}%
  \newcommand*\fsize{\dimexpr\f@size pt\relax}%
  \newcommand*\lineheight[1]{\fontsize{\fsize}{#1\fsize}\selectfont}%
  \ifx\svgwidth\undefined%
    \setlength{\unitlength}{459.89137184bp}%
    \ifx\svgscale\undefined%
      \relax%
    \else%
      \setlength{\unitlength}{\unitlength * \real{\svgscale}}%
    \fi%
  \else%
    \setlength{\unitlength}{\svgwidth}%
  \fi%
  \global\let\svgwidth\undefined%
  \global\let\svgscale\undefined%
  \makeatother%
  \begin{picture}(1,0.51917183)%
    \lineheight{1}%
    \setlength\tabcolsep{0pt}%
    \put(0,0){\includegraphics[width=\unitlength,page=1]{unconstrained_eq.pdf}}%
    \put(0.06989363,0.03183898){\makebox(0,0)[lt]{\lineheight{1.25}\smash{\begin{tabular}[t]{l}0\end{tabular}}}}%
    \put(0.32740946,0.02857734){\makebox(0,0)[lt]{\lineheight{1.25}\smash{\begin{tabular}[t]{l}1\end{tabular}}}}%
    \put(0.92663109,0.03001731){\makebox(0,0)[lt]{\lineheight{1.25}\smash{\begin{tabular}[t]{l}$\infty$\end{tabular}}}}%
    \put(0,0){\includegraphics[width=\unitlength,page=2]{unconstrained_eq.pdf}}%
    \put(0.15328085,0.4504947){\makebox(0,0)[lt]{\lineheight{1.25}\smash{\begin{tabular}[t]{l}\large $L(c)$\end{tabular}}}}%
    \put(0.10465262,0.16342969){\makebox(0,0)[lt]{\lineheight{1.25}\smash{\begin{tabular}[t]{l}\large $C(l)$\end{tabular}}}}%
    \put(0.65457803,0.00438595){\makebox(0,0)[lt]{\lineheight{1.25}\smash{\begin{tabular}[t]{l}\large $l$\end{tabular}}}}%
    \put(-0.00044381,0.28895279){\makebox(0,0)[lt]{\lineheight{1.25}\smash{\begin{tabular}[t]{l}\large $c$\end{tabular}}}}%
    \put(0,0){\includegraphics[width=\unitlength,page=3]{unconstrained_eq.pdf}}%
  \end{picture}%
\endgroup%
}
    \caption{A parameterization of the model with three unconstrained equilibria.}
    \label{fig: unconstrained}
\end{figure}

These properties imply the graphs of $C$ and $L$ must intersect, possibly multiple times, which means unconstrained equilibria exist. See Figure \ref{fig: unconstrained}.

Suppose groups $A$ and $B$ play distinct unconstrained equilibria, $(\ol{c_A^*}, \ul{l_A^*})$ and $(\ol{c_B^*}, \ul{l_B^*})$, respectively. Then $c_B^* \neq c_A^*$. Without loss of generality, assume $c_B^* < c_A^*$. Then the DM has worse beliefs about quality in group $B$ -- $f(x)\vert H(c_B^*) < f(x)\vert H(c_A^*)$ for all features $x$. This causes the DM to apply a tougher decision policy to group $B$ -- $l_B^* = L(c_B^*) > L(c_A^*) = l_A^*$. Group $B$ individuals then have less incentive to become qualified -- $Inc(\ul{l_B^*}) = C(l_B^*) = c_B^* < c_A^* = C(l_A^*) = Inc(\ul{l_A^*})$. This results in a worse distribution of quality throughout the group -- $H(c_B^*) < H(c_A^*)$, which ultimately fulfills the DM's worse beliefs about quality in the group.

Since $l_B^* > l_A^*$, a group $B$ individual has a strictly lower expected payoff than a group $A$ individual with the same cost. Together, $l_B^* > l_A^*$ and $H(c_B^*) < H(c_A^*)$ imply $V(\ul{l_B^*}, \mu\vert H(c_B^*), f\vert H(c_B^*)) < V(\ul{l_A^*}, \mu\vert H(c_A^*), f\vert H(c_A^*))$. Thus, $(\ol{c_B^*}, \ul{l_B^*})$ is Pareto-dominated by $(\ol{c_A^*}, \ul{l_A^*})$, and unconstrained equilibria are Pareto-ranked.

Given any two intersections of the graphs of $C$ and $L$, one is higher (and further to the left) than the other. The unconstrained equilibrium corresponding to the higher intersection Pareto-dominates the one corresponding to the lower intersection.

\begin{definition}
We say the DM statistically discriminates against group $B$ ($A$) if the two groups play distinct unconstrained equilibria, with group $A$ ($B$) playing the Pareto-superior one.
\end{definition}

\section{Interventions}

\begin{definition}
Let $\{(d_A, d_B)\}$ denote the set of all decision policy pairs, where $d_A$ and $d_B$ are applied to groups $A$ and $B$, respectively. An intervention $k$ maps the feature densities of the two groups, $(\mu_A, \mu_B)$, and the DM's beliefs about quality in the two groups, $(f_A, f_B)$, to a set of allowed decision policy pairs, $k(\mu_A, \mu_B, f_A, f_B) \subset \{(d_A, d_B)\}$.
\end{definition}

Interventions can fully condition on the DM's beliefs about quality. This reflects our assumption that the DM uses a machine learning algorithm to verifiably output $(f_A, f_B)$. By contrast, interventions cannot condition on the preference parameters of the model, $(H, v_q, v_u)$, as policymakers often do not know much about preferences.

Interventions also cannot condition on the data-generating parameters of the model, $(p_q, p_u)$. At first, this seems incongruent with our assumption that interventions can condition on $(f_A, f_B)$. Machine learning algorithms are trained on datasets, consisting of past individuals, each summarized by their group, features, and revealed quality. If the training dataset can be used to learn the conditional distribution of quality given features, as summarized by $(f_A, f_B)$, why can't it be used to learn $(p_q, p_u)$, the conditional distribution of features given quality?

The asymmetry has to do with the fact that, while quality is binary, feature vectors are often high-dimensional. For example, Upstart, an AI lending company, uses more than 1600 features to predict default \citep{upstart2026}. Training datasets are often rich enough to separate two qualities well, yielding an accurate $(f_A, f_B)$. But most training datasets are not rich enough to accurately estimate high-dimensional probability densities, such as $p_q$ and $p_u$ when $N \gg 1$. As a result, decision makers typically follow Vapnik's Principle, which calls for directly estimating the beliefs about quality that enter the utility function, rather than first estimating the data-generating parameters and then backing out the beliefs about quality using Bayes' rule.

\begin{definition}
Given an intervention $k$, an equilibrium between the two groups and a $k$-constrained DM -- or simply a $k$-constrained equilibrium -- consists of, for each group $g$, a quality profile $\ol{c_g^*}$, beliefs about quality $f_g^*$, and a decision policy $d_g^*$, satisfying the following conditions:
\begin{itemize}
  \item Each quality profile is the best-response to the corresponding decision policy, $(c_A^*, c_B^*) = (Inc(d_A^*), Inc(d_B^*))$.
  \item The beliefs about quality in each group are rational given the fraction of qualified individuals in that group, $(f_A^*, f_B^*) \equiv (f\vert H(c_A^*), f\vert H(c_B^*))$.
  \item The decision policy pair is a best-response given the intervention, the feature densities, and the beliefs about quality,
  \begin{align*}
  (d^*_A, d^*_B)		& \in \argmax_{(d_A, d_B) \in k\left(\mu\vert H(c_A^*), \mu\vert H(c_B^*), f_A^*, f_B^*\right)} \sum_{g \in \{A, B\}} \l_g V(d_g, \mu\vert H(c_g^*), f_g^*).
  \end{align*}
\end{itemize}
\end{definition}

\subsection{Common Identity}

Statistical discrimination involves the two groups playing distinct unconstrained equilibria. This motivates us to search for an intervention $k$ with the property that $k$-constrained equilibria coincide with scenarios in which both groups play the same unconstrained equilibrium.

Initially, there seems to be an obvious intervention with our desired property: A decision policy pair is allowed if and only if both components are the same likelihood-cutoff decision policy.

Applying the same likelihood-cutoff decision policy to both groups provides both groups the same incentive. So in any equilibrium under the proposed intervention, both groups are equally qualified with the same quality profile. When both groups have the same quality profile, the DM's unconstrained best-response, given rational beliefs about quality, is to apply the same likelihood-cutoff decision policy to both groups. The proposed intervention then ceases to be a binding constraint on the DM and can be lifted. In this way, intervention-constrained equilibria coincide with scenarios in which both groups play the same unconstrained equilibrium.

The problem with the proposed intervention is that, whether or not a decision policy has a likelihood-cutoff structure depends on the data-generating parameters $(p_q, p_u)$. Sometimes, one can identify the set of likelihood-cutoff decision policies without knowing those parameters. For example, when the only feature is the score, $x \in [0, 1]$, on a test of quality, one knows the likelihood function $\frac{p_q(x)}{p_u(x)}$ is increasing in $x$, even if one does not know $(p_q, p_u)$. In this case, a decision policy has a likelihood-cutoff structure if and only if it is a cutoff function of $x$. 

However, once the feature space is multi-dimensional, there is no longer any natural direction in which the likelihood function must increase. Now, it seems, identifying the set of likelihood-cutoff decision policies requires knowing $(p_q, p_u)$, which interventions cannot condition on.

The key insight of this paper is:

\begin{quote}
\textit{The DM's beliefs about quality in a group can be used to identify the set of likelihood-cutoff decision policies.}
\end{quote}

To see how, let $\pi_g \in (0, 1)$ be the fraction of qualified individuals in group $g$, and let the DM's beliefs about quality be rational. Since
\begin{align*}
f(x) \vert \pi_g := \frac{\pi_g p_q(x)}{\pi_g p_q(x) + (1 - \pi_g)p_u(x)} = \frac{\pi_g \cdot \frac{p_q(x)}{p_u(x)}}{\pi_g \cdot \frac{p_q(x)}{p_u(x)} + (1 - \pi_g)},
\end{align*}
the DM's beliefs about quality in group $g$ are comonotonic with the likelihood function,
\begin{align*}
f(x)\vert \pi_g < f(x')\vert \pi_g\ \Leftrightarrow\ \frac{p_q(x)}{p_u(x)} < \frac{p_q(x')}{p_u(x')} \ \ \ \forall x, x' \in [0, 1]^N.
\end{align*}

More generally, let $(\a_A, \a_B)$ be a probability weighting, meaning $\a_A + \a_B = 1$ and $\a_A, \a_B \geq 0$. Then the weighted average, $\a_A f\vert \pi_A + \a_B f\vert \pi_B$, is comonotonic with the likelihood function. Comonotonicity implies that a decision policy has a likelihood-cutoff structure if and only if it is a cutoff function of $\a_A f\vert \pi_A + \a_B f\vert \pi_B$. We now have our desired intervention:

\begin{definition}
Fix a probability weighting $(\a_A, \a_B)$. The common identity intervention allows the set $k_{CI}(\mu_A, \mu_B, f_A, f_B) =$
\begin{align*}
\big\{(d_A, d_B)\ \vert\ \mbox{$d_A$ and $d_B$ are the same cutoff function of $\a_A f_A + \a_B f_B$}\big\}.
\end{align*}
A decision policy $d$ is a cutoff function of $\a_A f_A + \a_B f_B$ if there exists an $s \in [0, 1]$ such that $d(x) = 1$ if and only if $\a_A f_A(x) + \a_B f_B(x) \geq s$.
\end{definition}

Common identity is a family of interventions, parameterized by the probability weighting $(\a_A, \a_B)$. The one described in the Introduction has parameter $(1, 0)$. The population's group distribution, $(\l_A, \l_B)$, is another possibility. When the training dataset is representative of the population, the $(\l_A, \l_B)$ parameterization can minimize estimation error. However, if one group suffers from small sample size, then it may be better to overweight the other group. Also, \citet{gillis2024orthogonalizing} shows that, when the training procedure is LASSO, inclusion of group membership as an explanatory variable is unstable. In this case, it is best to set $\a_A$ or $\a_B$ to zero \citep{yang2020equal}.

I abstract from issues with the training dataset or procedure that cause one parameterization of common identity to be more practical than another. My results apply to all parameterizations of common identity.

The analysis leading up to the definition of common identity proves
\begin{lemma}\label{obj_1}
Suppose the fraction of qualified individuals in each group is strictly between zero and one, and the DM's beliefs about quality are rational. Then a decision policy pair is allowed by common identity if and only if both components are the same likelihood-cutoff decision policy.
\end{lemma}

Let $\{(\ol{c_A^*}, d_A^*), (\ol{c_B^*}, d_B^*)\}$ be a common-identity-constrained equilibrium -- there is no need to specify the DM's rational beliefs about quality, which are determined by the quality profiles. Since $H(c_A^*), H(c_B^*) \in (0, 1)$, Lemma \ref{obj_1} implies there is an $\ul{l^*}$ such that $d_A^* \equiv d_B^* \equiv \ul{l^*}$, which then implies $c_A^* = c_B^* = C(l^*)$. Let $c^* := C(l^*)$. Since both groups have quality profile $\ol{c^*}$, the DM's unconstrained best-response is $(\ul{L(c^*)}, \ul{L(c^*)})$. By Lemma \ref{obj_1}, $(\ul{L(c^*)}, \ul{L(c^*)})$ is allowed by common identity. So $l^* = L(c^*)$ and $(\ol{c^*}, \ul{l^*})$ is an unconstrained equilibrium.

Conversely, let $(\ol{c^*}, \ul{l^*})$ be an unconstrained equilibrium. Since $H(c^*) \in (0, 1)$, by Lemma \ref{obj_1}, when both groups have quality profile $\ol{c^*}$ and the DM has rational beliefs about quality, $(\ul{l^*}, \ul{l^*})$ is allowed by common identity. So $\{(\ol{c^*}, \ul{l^*}), (\ol{c^*}, \ul{l^*})\}$ is a common-identity-constrained equilibrium. We have now proved

\begin{theorem}\label{obj_2}
A strategy profile $\{(\ol{c_A^*}, d_A^*), (\ol{c_B^*}, d_B^*)\}$ is a common-identity-constrained equilibrium if and only if there is an unconstrained equilibrium $(\ol{c^*}, \ul{l^*})$ such that $(\ol{c_A^*}, d_A^*) \equiv (\ol{c_B^*}, d_B^*) \equiv (\ol{c^*}, \ul{l^*})$.
\end{theorem}

Theorem \ref{obj_2} states that common-identity-constrained equilibria coincide with scenarios in which both groups play the same unconstrained equilibrium. The characterization begs a few questions.

First, given that unconstrained equilibria are Pareto-ranked, why not go for an intervention with a unique intervention-constrained equilibrium that coincides with both groups playing the Pareto-dominant unconstrained equilibrium? Unfortunately, two parameterizations of the model can share an unconstrained equilibrium, that is Pareto-dominant in one parameterization and not Pareto-dominant in the other. This means even identifying the Pareto-dominant unconstrained equilibrium requires conditioning on model parameters, which our interventions cannot do.

Second, why require intervention-constrained equilibrium play to coincide with unconstrained equilibrium play at all? A binding intervention compels the DM to take suboptimal actions. This distorts market forces and can lead to unintended consequences elsewhere in the economy. A binding intervention also requires a mechanism to enforce compliance by the DM, which can be expensive to maintain. Therefore, it is natural to want an intervention that eventually renders itself unnecessary.

Finally, which unconstrained equilibrium emerges when common identity is imposed? Suppose the DM statistically discriminates against group $B$. After imposing common identity, we would obviously rather group $B$ converge to group $A$'s Pareto-superior unconstrained equilibrium than the other way around. Let us now provide conditions for when such a convergence happens under best-response dynamics. However, we leave a full-fledged dynamic theory of interventions against statistical discrimination to future work.

\subsubsection{Best-Response Dynamics}\label{best-response}

Imagine the DM initially applies a likelihood-cutoff decision policy $\ul{l_0}$ to a group, and then the group and the DM alternately best-respond to each other's current strategy. The resulting sequence of decision policies and quality profiles is
\begin{align}\label{adaptive}
\ul{l_0}, \ol{c_0} := \ol{C(l_0)}, \ul{l_1} := \ul{L(c_0)}, \ol{c_1} := \ol{C(l_1)}, \ul{l_2} := \ul{L(c_1)} \ldots,
\end{align}
which we call the \emph{best-response dynamics starting from $\ul{l_0}$}.

\begin{definition}
The basin of attraction for an unconstrained equilibrium $(\ol{c^*}, \ul{l^*})$ is the set of $l_0$ such that the best-response dynamics starting from $\ul{l_0}$ satisfy $\lim_{t \rightarrow \infty} l_t = l^*$.
\end{definition}

Let $(\ol{c^*}, \ul{l^*})$ be an unconstrained equilibrium and $l_0$ be in its basin of attraction. Since $C$ is continuous and $c^* = C(l^*)$, the best-response dynamics starting from $\ul{l_0}$ also satisfy $\lim_{t \rightarrow \infty} c_t = c^*$. So, if the DM initially applies $\ul{l_0}$ to a group, then under best-response dynamics, the group and the DM will converge to playing $(\ol{c^*}, \ul{l^*})$.

\begin{definition}
An unconstrained equilibrium $(\ol{c^*}, \ul{l^*})$ is locally stable if its basin of attraction contains an open interval around $l^*$.
\end{definition}

Locally stable unconstrained equilibria exist under fairly general conditions: Suppose $C$ and $L$ are differentiable, and their graphs intersect more than once, so that statistical discrimination is possible. Suppose the intersections are never tangential. Then the number of intersections is odd and at least three. Counting them from left to right, all intersections, except possibly the first one, occur in the region $l > 1$. If $(c^*, l^*)$ is an odd numbered intersection satisfying $l^* > 1$, then the graph of $L$ is steeper than the graph of $C$ at $(c^*, l^*)$ -- i.e., $\vert \frac{1}{L'(c^*)} \vert > \vert C'(l^*)\vert$. As pointed out by \citet{coate1993will}, an unconstrained equilibrium is locally stable whenever the corresponding intersection of the graphs of $C$ and $L$ has this property.

\begin{figure}[t]
  \centering
    \centering
    \scalebox{.8}{%
 \executeiffilenewer{InkscapePics/unconstrained_adaptive.svg}{InkscapePics/unconstrained_adaptive.pdf}%
 {inkscape -z -D --file=InkscapePics/unconstrained_adaptive.svg %
 --export-pdf=InkscapePics/unconstrained_adaptive.pdf --export-latex}%
 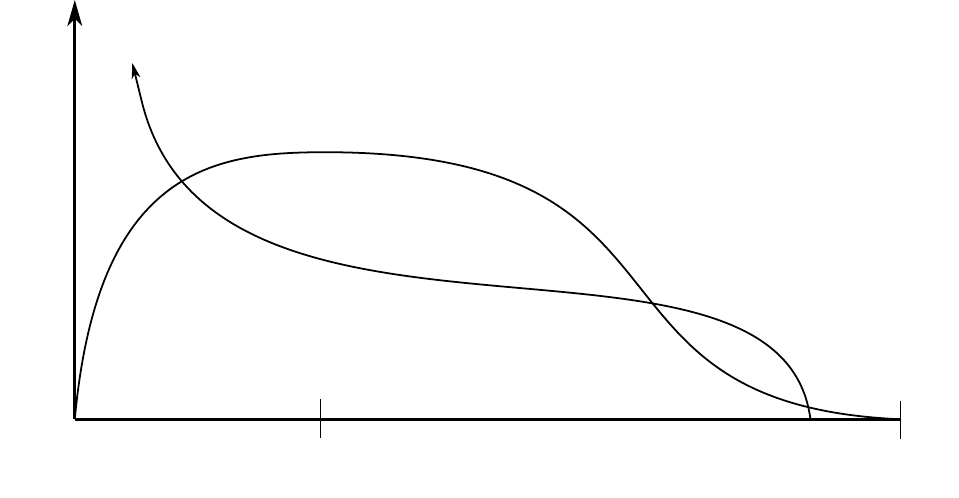%
}
    \caption{Some best-response dynamics and local stability.}
    \label{fig: unconstrained_adaptive}
\end{figure}

For example, Figure \ref{fig: unconstrained_adaptive} depicts a parameterization of the model such that the graphs of $C$ and $L$ intersect three times. At every odd intersection, the graph of $L$ is steeper than the graph of $C$. Using Figure \ref{fig: unconstrained_adaptive}, one can directly check that the basin of attraction for $(\ol{c_3^*}, \ul{l_3^*})$ contains $(l_2^*, \infty)$, which is an open interval around $l_3^*$. So, $(\ol{c_3^*}, \ul{l_3^*})$ is indeed locally stable. The arrows diverging from near $(c_2^*, l_2^*)$ in Figure \ref{fig: unconstrained_adaptive} imply that $(\ol{c_2^*}, \ul{l_2^*})$ is not locally stable. Finally, Figure \ref{fig: unconstrained_adaptive} shows that $L(C(L(C(1)))) < 1$, which implies the basin of attraction for $(\ol{c_1^*}, \ul{l_1^*})$ contains $(L(C(1)), l_2^*)$, an open interval around $l_1^*$. So $(\ol{c_1^*}, \ul{l_1^*})$ is also locally stable.

\begin{theorem}\label{converge}
Let $(\ol{c_A^*}, \ul{l_A^*})$ and $(\ol{c_B^*}, \ul{l_B^*})$ be distinct unconstrained equilibria, and let
\begin{align*}
l_0 := \argmax_{l \in [0, \infty]} \l_A V(\ul{l}, \mu\vert H(c_A^*), f\vert H(c_A^*)) + \l_B V(\ul{l}, \mu\vert H(c_B^*), f\vert H(c_B^*)).
\end{align*}
If $(\ol{c_g^*}, \ul{l_g^*})$ is locally stable and $\l_g$ is sufficiently close to 1, then the best-response dynamics starting from $\ul{l_0}$ satisfy $\lim_{t \rightarrow \infty} (c_t, l_t) = (c_g^*, l_g^*)$.
\end{theorem}

Note: It is well-defined to fix an unconstrained equilibrium and, more specifically, a locally stable unconstrained equilibrium, while varying $(\l_A, \l_B)$, because the set of unconstrained equilibria and local stability do not depend on $(\l_A, \l_B)$. 

The $l_0$ defined in Theorem \ref{converge} lies between $l_A^*$ and $l_B^*$. Moreover, the larger $\l_g$ is, the closer $l_0$ is to $l_g^*$ with $l_0$ converging to $l_g^*$ as $\l_g$ converges to 1. This means, given any open interval around $l_g^*$, if $\l_g$ is sufficiently close to 1, then $l_0$ lies in that interval. The definition of local stability then implies Theorem \ref{converge}.

How should Theorem \ref{converge} be interpreted? Imagine groups $A$ and $B$ initially play $(\ol{c_A^*}, \ul{l_A^*})$ and $(\ol{c_B^*}, \ul{l_B^*})$, respectively. Imposing common identity causes the DM to deviate from $(\ul{l_A^*}, \ul{l_B^*})$ to $(\ul{l_0}, \ul{l_0})$. Both groups then best-respond with the same quality profile, $\ol{c_0}$ from \eqref{adaptive}. Once both groups have the same quality profile, common identity ceases to be a binding constraint and can be lifted. Theorem \ref{converge} then implies, if $(\ol{c_g^*}, \ul{l_g^*})$ is locally stable and $\l_g$ is sufficiently close to 1, each group and the DM converge, under best-response dynamics, to playing $(\ol{c_g^*}, \ul{l_g^*})$. In particular, 

\begin{remark}
Suppose the DM statistically discriminates against group $B$. If group $A$'s unconstrained equilibrium is locally stable and the fraction of the population belonging to group $A$ is sufficiently large, then a temporary program of common identity causes each group and the DM to converge, under best-response dynamics, to playing group $A$'s Pareto-superior unconstrained equilibrium.
\end{remark}

Revisiting the example in Figure \ref{fig: unconstrained_adaptive}, let group $A$ play the Pareto-dominant $(\ol{c_1^*}, \ul{l_1^*})$, group $B$ play the Pareto-worst $(\ol{c_3^*}, \ul{l_3^*})$, and $\l_A$ be large enough to ensure $\argmax_{l \in [0, \infty]}$ $\l_A V(\ul{l}, \mu\vert H(c_1^*), f\vert H(c_1^*)) + \l_B V(\ul{l}, \mu\vert H(c_3^*), f\vert H(c_3^*)) < l_2^*$. Then a temporary program of common identity causes each group and the DM to converge, under best-response dynamics, to playing the Pareto-dominant $(\ol{c_1^*}, \ul{l_1^*})$.

\subsection{Features that Depend on Group Membership}\label{group_depend}

The original model assumes that features depend on quality but not on group membership. Let us now generalize the original model in two distinct ways that allow features to also depend on group membership. In the first way, features depend on quality and group membership separably -- some features depend only on quality and, conditional on those features, the rest depend only on group membership. In the second way, features jointly depend on quality and group membership.

\subsubsection{Separable Dependence on Quality and Group Membership}\label{separable}

In this section, individuals have features consisting of two components, $x = (x_Q, x_G)$. Here, $x_Q$ depends only on quality, while $x_G$ depends on group membership and is conditionally independent of quality given $x_Q$. For example, consider a hiring setting where the unobserved quality is productivity, and an individual's $x_Q$ includes data on income. Think about the individual's zip code. It is plausible that zip code depends on productivity only through income, which determines the neighborhoods the individual can afford. In addition, zip code may be influenced by group-specific preferences. In this case, zip code is part of $x_G$.

We now generalize the original model to allow for such ``separable" dependence of features on quality and group membership. To create the separable model, begin by fixing a parameterization of the original model -- call it the \emph{underlying original model.} Let $Q$ denote the underlying original model's set of $N$ feature types and relabel its feature space as $[0, 1]^Q$. Refer to elements of $[0, 1]^Q$ by $x_Q$. Next, introduce an additional (possibly empty) set of feature types $G$. A vector of $G$ features is denoted $x_G$ and takes values in $[0, 1]^G$. For each $x_Q \in [0, 1]^Q$, introduce full-support densities $p_A(x_G \vert x_Q)$ and $p_B(x_G \vert x_Q)$ over $[0, 1]^G$. For each group $g$, the densities of a vector of features $x = (x_Q, x_G)$, conditional on being qualified and unqualified, are $p_q(x_Q)p_g(x_G \vert x_Q)$ and $p_u(x_Q)p_g(x_G \vert x_Q)$, respectively. The separable model is now fully parameterized, with feature space $[0, 1]^Q \times [0, 1]^G$. When $G$ is empty, the separable model reduces to the underlying original model.

Intuitively, since $x_G$ is uninformative about quality conditional on $x_Q$, an unconstrained DM with rational beliefs about quality ignores $x_G$. Unconstrained equilibria in the separable model should then be equivalent to those in the underlying original model.

To prove the conjecture, first note that, in the separable model, the likelihood function depends on $x$ only up to $x_Q$, $\frac{p_q(x_Q)p_g(x_G \vert x_Q)}{p_u(x_Q)p_g(x_G \vert x_Q)} = \frac{p_q(x_Q)}{p_u(x_Q)}$. Viewed as a function of $x_Q$, it is the same as the likelihood function of the underlying original model.

Let $\pi_g$ be the fraction of qualified individuals in group $g$. In the separable model, the feature densities of the two groups can differ, even if both groups have the same fraction of qualified individuals. To emphasize this, we denote group $g$'s feature density by $\mu\vert_g \pi_g$, instead of $\mu \vert \pi_g$, where $\mu(x)\vert_g \pi_g := \pi_g p_q(x_Q)p_g(x_G \vert x_Q) + (1-\pi_g) p_u(x_Q)p_g(x_G \vert x_Q)$. However, the rational beliefs about quality in a group still depend only on the fraction of qualified individuals in the group. For group $g$, they are $f\vert \pi_g$, where
  \begin{align*}
 f(x)\vert \pi_g := \frac{\pi_g p_q(x_Q)}{\pi_g p_q(x_Q) + (1-\pi_g) p_u(x_Q)}.
  \end{align*}
Viewed as a function of $x_Q$, the rational beliefs about quality in a group are the same as those in the underlying original model. Thus, when a group's quality profile is $\ol{c}$, the DM's best-response, given rational beliefs about quality in the group, is still to accept a member with features $x$ if and only if the likelihood of $x$ is at least $\left[ \frac{1 - H(c)}{H(c)}\right] \frac{v_u}{v_q}$, which is the $L(c)$ of the underlying original model.

In the separable model, if a decision policy $d$ depends on $x$ only up to $x_Q$, then the incentive to become qualified facing $d$ remains group-independent: Group $A$'s incentive to become qualified facing such a $d$ is
\begin{align*}
 			& = \int_{[0, 1]^N} d(x) \left( p_q(x_Q)p_A(x_G \vert x_Q) - p_u(x_Q)p_A(x_G \vert x_Q) \right) dx\\
			& = \int_{[0, 1]^Q} d(x_Q) \left( p_q(x_Q) - p_u(x_Q) \right) dx_Q\\
 			& = \int_{[0, 1]^N} d(x) \left( p_q(x_Q)p_B(x_G \vert x_Q) - p_u(x_Q)p_B(x_G \vert x_Q) \right) dx,
\end{align*}
which is group $B$'s incentive to become qualified facing $d$. In particular, the incentive to become qualified facing a likelihood-cutoff decision policy $\ul{l}$ remains group-independent, and equals the $C(l)$ of the underlying original model.

So, unconstrained equilibria still correspond to intersections of the graphs of the underlying original model's $C$ and $L$.

\begin{lemma}
A quality profile $\ol{c^*}$ and a likelihood-cutoff decision policy $\ul{l^*}$ comprise an unconstrained equilibrium in the separable model if and only if they comprise an unconstrained equilibrium in the underlying original model.
\end{lemma}

The likelihood function and the DM's rational beliefs about quality in each group, viewed as functions of $x_Q$, are the same as in the underlying original model. So, by the same reasoning as in the original model, Lemma \ref{obj_1} remains true. Also, by the same reasoning as in the original model, Lemma \ref{obj_1} implies that common-identity-constrained equilibria coincide with scenarios in which both groups play the same unconstrained equilibrium:

\begin{theorem}\label{separable_CI}
In the separable model, a strategy profile $\{(\ol{c_A^*}, d_A^*), (\ol{c_B^*}, d_B^*)\}$ is a common-identity-constrained equilibrium if and only if there is an unconstrained equilibrium $(\ol{c^*}, \ul{l^*})$ such that $(\ol{c_A^*}, d_A^*) \equiv (\ol{c_B^*}, d_B^*) \equiv (\ol{c^*}, \ul{l^*})$.
\end{theorem}

Let us now compare the performance of common identity to that of group-blinding.

\begin{definition}
The group-blinding intervention allows the DM to choose any decision policy if it is applied to both groups: $k_{blind}(\mu_A, \mu_B, f_A, f_B) = \{(d_A, d_B)\ \vert\ d_A \equiv d_B\}$.
\end{definition}

In the original model, group-blinding performs just as well as common identity. 

\begin{proposition}\label{group_blind}
In the separable model, when $G$ is empty, a strategy profile $\{(\ol{c_A^*}, d_A^*)$, $(\ol{c_B^*}, d_B^*)\}$ is a group-blinding-constrained equilibrium if and only if there is an unconstrained equilibrium $(\ol{c^*}, \ul{l^*})$ such that $(\ol{c_A^*}, d_A^*) \equiv (\ol{c_B^*}, d_B^*) \equiv (\ol{c^*}, \ul{l^*})$.
\end{proposition}
\begin{proof}
See the appendix.
\end{proof}

However, once features depend on group membership, group-blinding's performance degrades, making common identity an attractive alternative. 

\begin{proposition}\label{group_blind_separable}
There exist parameterizations of the separable model that have group-blinding-constrained equilibria involving unequally qualified groups -- that is, one group has a smaller fraction of qualified individuals than the other group.
\end{proposition}

\begin{figure}[t]
  \centering
    \centering
    \scalebox{.8}{%
 \executeiffilenewer{InkscapePics/unconstrained_blind.svg}{InkscapePics/unconstrained_blind.pdf}%
 {inkscape -z -D --file=InkscapePics/unconstrained_blind.svg %
 --export-pdf=InkscapePics/unconstrained_blind.pdf --export-latex}%
 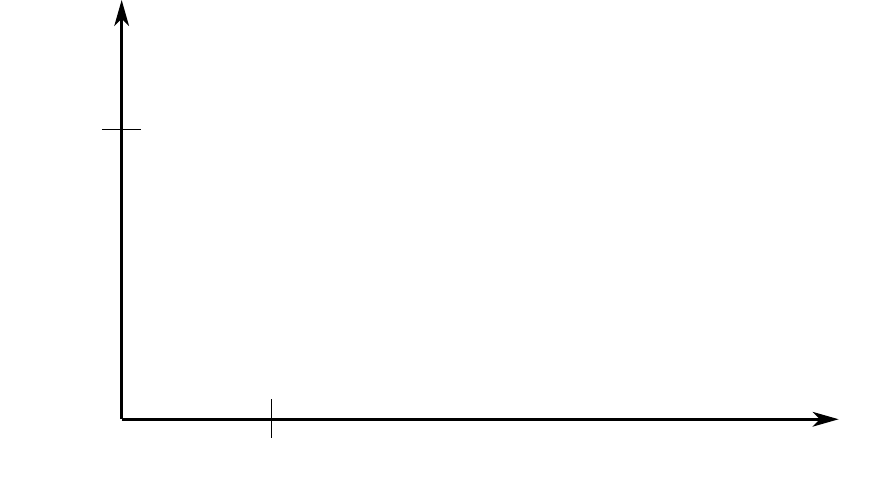%
}
    \caption{A group-blinding-constrained equilibrium involving unequally qualified groups.}
    \label{fig: unconstrained_blind}
\end{figure}

The proof of Proposition \ref{group_blind_separable} is constructive: Fix an underlying original model with $C$ and $L$ as in Figure \ref{fig: unconstrained_blind}. In this model, $(\ol{c_A^*}, \ul{l_A^*})$ and $(\ol{c_B^*}, \ul{l_B^*})$ are two (locally stable) unconstrained equilibria, satisfying $1 < l_A^* < l_B^*$, with a (not locally stable) unconstrained equilibrium $(\ol{c^*}, \ul{l^*})$ in between. To complete the parameterization of the separable model, let $[0, 1]^G := [0, 1]$, $\hat{A} := [0, \frac{1}{2}]$, and $\hat{B} := (\frac{1}{2}, 1]$. Fix $\e \in (0, \frac{1}{2})$ and assume $p_g(x_G \vert x_Q) = p_g(x_G)$ for all $(x_Q, x_G)$ and both groups $g$, where
\begin{align*}
p_A(x_G) =
\begin{cases}
2 - 2\e	& \mbox{if $x_G \in \hat{A}$}\\
2\e		& \mbox{if $x_G \in \hat{B}$}
\end{cases}
& &p_B(x_G) =
\begin{cases}
2\e		& \mbox{if $x_G \in \hat{A}$}\\
2 - 2\e	& \mbox{if $x_G \in \hat{B}$}
\end{cases}
\end{align*}
Intuitively, $x_G \in \hat{A}$ ($\hat{B}$) is a noisy signal that the individual belongs to group $A$ ($B$). 

The decision policy pair, $(\ul{l_A^*}, \ul{l_B^*})$, which supports statistical discrimination against group $B$, is not allowed under group-blinding. Following the discussion in the Introduction, a group-blinding-constrained DM can try to mimic $(\ul{l_A^*}, \ul{l_B^*})$ by using $x_G$ as a proxy for group membership. Specifically, the DM selects $l_{\hat{A}} \in (l_A^*, l^*)$ and $l_{\hat{B}} \in (l^*, l_B^*)$ -- akin to the two test score cutoffs described in the Introduction. Then, whenever an individual's $x_G \in \hat{A}$ ($\hat{B}$), the DM accepts the individual if and only if the likelihood of the individual's $x_Q$ is at least $l_{\hat{A}}$ ($l_{\hat{B}}$). Call this decision policy $\hat{d}$.

Facing $\hat{d}$, group $A$'s incentive to become qualified is
\begin{align*}
	& \int_{[0, 1]^G} \int_{[0, 1]^Q} \hat{d}(x_Q, x_G) (p_q(x_Q)p_A(x_G) - p_u(x_Q)p_A(x_G)) dx_Q dx_G\\
=	& (1-\e)  \int_{[0, 1]^Q} \ul{l_{\hat{A}}}(x_Q) (p_q(x_Q) - p_u(x_Q)) dx_Q + \e \int_{[0, 1]^Q} \ul{l_{\hat{B}}}(x_Q) (p_q(x_Q) - p_u(x_Q)) dx_Q\\
=	& (1-\e) C(l_{\hat{A}}) + \e C(l_{\hat{B}}).
\end{align*}
So, group $A$'s best-response quality profile is $\ol{c_{\hat{A}}}$, where $c_{\hat{A}} := (1-\e) C(l_{\hat{A}}) + \e C(l_{\hat{B}})$. Similarly, group $B$'s best-response quality profile is $\ol{c_{\hat{B}}}$, where $c_{\hat{B}} := \e C(l_{\hat{A}}) + (1-\e) C(l_{\hat{B}})$. Since $\e < \frac{1}{2}$ and $C(l_{\hat{A}}) > C(l_{\hat{B}})$, we have $c_{\hat{A}} > c_{\hat{B}}$, which means, when the two groups best-respond to $\hat{d}$, group $B$ is less qualified than group $A$.

Under group-blinding, the DM must accept all individuals with features $x$ or reject all of them. When each group $g$ has quality profile $\ol{c_{\hat{g}}}$, the DM, given rational beliefs about quality in the two groups, is better off accepting all individuals with features $x = (x_Q, x_G)$ if and only if
\begin{align}\label{accept_all_x}
			& \sum_{g \in \{A, B\}} \l_g (\mu(x) \vert H(c_{\hat{g}})) \left[ \left(f(x)\vert H(c_{\hat{g}})\right)v_q - \left(1- f(x)\vert H(c_{\hat{g}})\right) v_u\right] \geq 0 \notag\\
\Leftrightarrow 	& \sum_{g \in \{A, B\}} \l_g p_g(x_G) \left[ H(c_{\hat{g}}) p_q(x_Q) v_q - (1 - H(c_{\hat{g}})) p_u(x_Q)v_u\right] \geq 0.
\end{align}
When $x_G \in \hat{A}$, Inequality \eqref{accept_all_x} is equivalent to
\begin{gather*}
\l_A (1-\e)\left[ H(c_{\hat{A}}) \frac{p_q(x_Q)}{p_u(x_Q)} - (1- H(c_{\hat{A}}))\frac{v_u}{v_q}\right] \geq - \l_B \e\left[ H(c_{\hat{B}}) \frac{p_q(x_Q)}{p_u(x_Q)} - (1- H(c_{\hat{B}}))\frac{v_u}{v_q}\right]\\
\Leftrightarrow \frac{p_q(x_Q)}{p_u(x_Q)} \geq \left[\frac{(\l_A(1-\e) + \l_B \e) - (\l_A (1-\e)H(c_{\hat{A}}) + \l_B \e H(c_{\hat{B}}))}{\l_A (1-\e) H(c_{\hat{A}}) + \l_B \e H(c_{\hat{B}})}\right]  \frac{v_u}{v_q}\\
\Leftrightarrow \frac{p_q(x_Q)}{p_u(x_Q)} \geq L(CE_A(c_{\hat{A}}, c_{\hat{B}})),
\end{gather*}
where $CE_A(c_{\hat{A}}, c_{\hat{B}}) := H^{-1}\left[ \frac{\l_A (1-\e)}{\l_A(1-\e) + \l_B \e} H(c_{\hat{A}}) + \frac{\l_B \e}{\l_A(1-\e) + \l_B \e} H(c_{\hat{B}})\right]$. One can think of $CE_A(c_{\hat{A}}, c_{\hat{B}})$ as the certainty equivalent of the lottery $\left(\frac{\l_A (1-\e)}{\l_A(1-\e) + \l_B \e}, \frac{\l_B \e}{\l_A(1-\e) + \l_B \e}\right)$ over $(c_{\hat{A}}, c_{\hat{B}})$ with respect to $H$ viewed as the utility function. Similarly, when $x_G \in \hat{B}$, Inequality \eqref{accept_all_x} is equivalent to $\frac{p_q(x_Q)}{p_u(x_Q)} \geq L(CE_B(c_{\hat{A}}, c_{\hat{B}}))$, where $CE_B(c_{\hat{A}}, c_{\hat{B}}) := H^{-1}\left[ \frac{\l_A \e}{\l_A \e + \l_B (1-\e)} H(c_{\hat{A}}) + \frac{\l_B (1-\e)}{\l_A \e + \l_B (1-\e)} H(c_{\hat{B}})\right]$. Since $c_{\hat{A}}$ and $c_{\hat{B}}$ depend on $(l_{\hat{A}}, l_{\hat{B}})$, we can view $CE_A$ and $CE_B$ as depending on $(l_{\hat{A}}, l_{\hat{B}})$.

Thus, $\{(\ol{c_{\hat{A}}}, \hat{d}), (\ol{c_{\hat{B}}}, \hat{d})\}$ is a group-blinding-constrained equilibrium -- in which group $B$ is less qualified than group $A$ -- if and only if $(l_{\hat{A}}, l_{\hat{B}})$ satisfies $L(CE_A(l_{\hat{A}}, l_{\hat{B}})) = l_{\hat{A}}$ and $L(CE_B(l_{\hat{A}}, l_{\hat{B}})) = l_{\hat{B}}$. To complete the proof of Proposition \ref{group_blind_separable}, it suffices to prove that such an $(l_{\hat{A}}, l_{\hat{B}})$ exists, for all sufficiently small $\e \in (0, \frac{1}{2})$.

Fix arbitrary $l_A^+ \in (l_A^*, l^*)$ and $l_B^- \in (l^*, l_B^*)$. Let us view $CE_A$ and $CE_B$ as functions over $[l_A^*, l_A^+] \times [l_B^-, l_B^*]$. As $\e \rightarrow 0$, $CE_A(l_A^+, l_B^*)$ and $CE_B(l_A^*, l_B^-)$ converge to $C(l_A^+)$ and $C(l_B^-)$, respectively. Since $L(C(l_A^+)) < l_A^+$ and $L(C(l_B^-)) > l_B^-$, for all sufficiently small $\e \in (0, \frac{1}{2})$, $L(CE_A(l_A^+, l_B^*)) < l_A^+$ and $L(CE_B(l_A^*, l_B^-)) > l_B^-$. Fix such an $\e$.

If $\e$ were 0, $CE_A(l_A^*, l_B^-)$ would equal $C(l_A^*)$. Because the $\e$ we fixed is positive and $C(l_B^-) < C(l_A^*)$, $CE_A(l_A^*, l_B^-) < C(l_A^*)$. So $L(CE_A(l_A^*, l_B^-)) > L(C(l_A^*)) = l_A^*$. In addition, $L(CE_A(l_A, l_B))$ is continuously increasing in both $l_A$ and $l_B$. These two observations, combined with the earlier observation that $L(CE_A(l_A^+, l_B^*)) < l_A^+$, imply $L(CE_A(l_A, l_B))$ continuously maps $[l_A^*, l_A^+] \times [l_B^-, l_B^*]$ to $(l_A^*, l_A^+)$.  Similarly, $L(CE_B(l_A, l_B))$ continuously maps $[l_A^*, l_A^+] \times [l_B^-, l_B^*]$ to $(l_B^-, l_B^*)$. So, by Brouwer's Fixed Point Theorem, there exists an $(l_{\hat{A}}, l_{\hat{B}}) \in (l_A^*, l_A^+) \times (l_B^-, l_B^*)$ such that $L(CE_A(l_{\hat{A}}, l_{\hat{B}})) = l_{\hat{A}}$ and $L(CE_B(l_{\hat{A}}, l_{\hat{B}})) = l_{\hat{B}}$. Proposition \ref{group_blind_separable} is now proved.

\subsubsection{Feature Bias}\label{feature_bias}

Many training datasets include features that are meant to depend only on quality but, for various reasons, also end up depending on group membership. A test may be designed to measure intellect, but rural students may have less access to test-prep services than urban students. If so, then even after controlling for intellect, rural students will tend to score lower than urban students. Or, consider the number of prior arrests. While this feature is indicative of an individual's criminal proclivity, it is also affected by policing intensity. If blacks are policed more intensely than whites, then even after controlling for criminal proclivity, blacks will tend to have more prior arrests than whites.

In these examples, the inferences about quality that can be drawn from specific feature values would differ across groups, even if the groups were equally qualified. This is known as \emph{feature bias}. To understand how common identity should be modified to handle feature bias, let us begin by informally working through an example.

Let the feature space be $[0, 1]$ and think of $x \in [0, 1]$ as the score on a test of quality. For each group $g$, let $p_{g, q}(x)$ and $p_{g, u}(x)$ denote the group-specific full-support densities of $x$ conditional on being qualified and unqualified, respectively. Imagine that a ``rural" group $B$ individual, who scores $x$, would have scored $T(x)$ had they been a member of ``urban" group $A$. Specifically, there is a differentiable monotone bijection $T: [0, 1] \rightarrow [0, 1]$ such that the feature densities of the groups satisfy,
\begin{align}\label{feature_transform}
p_{B, q}(x) = p_{A, q}(T(x))T'(x) \mbox{\ \ \ and\ \ \ } p_{B, u}(x) = p_{A, u}(T(x))T'(x).
\end{align}
One can think of $T$ as the boost that urban students experience due to better access to test-prep services. 

Condition \eqref{feature_transform} implies that the likelihood function is group-specific, with the group $B$ likelihood of $x$ equal to the group $A$ likelihood of $T(x)$. Let us assume both likelihood functions are increasing in $x$, given the interpretation of $x$ as the score on a test of quality.

What happens if common identity is imposed? Suppose the DM's beliefs about quality in the two groups, $(f_A, f_B)$, are rational. For each group $g$, $f_g$ is comonotonic with group $g$'s likelihood function, which is increasing in $x$. So $\a_A f_A + \a_B f_B$ is increasing in $x$. Complying with common identity then reduces to applying the same test score cutoff for acceptance $s$ to both groups. Given that $\frac{p_{B, q}(s)}{p_{B, u}(s)} = \frac{p_{A, q}(T(s))}{p_{A, u}(T(s))}$, if $T(s) > s$ -- as one would expect if group $A$ has better access to test-prep services -- then applying the same test score cutoff for acceptance $s$ to both groups actually means group $B$ faces a higher likelihood cutoff for acceptance than group $A$. If $T(s) - s$ is sufficiently large, the likelihood cutoff gap between the two groups, under common identity, can be wider than when the DM statistically discriminates against group $B$. That feature bias can actually cause common identity to widen group disparities was first pointed out by \citet{o2024nature}.

To ensure the DM applies the same likelihood cutoff for acceptance to both groups, common identity must be modified in the following way: A decision policy pair $(d_A, d_B)$ is allowed if and only if $d_A \circ T$ (instead of $d_A$) and $d_B$ are the same cutoff function of $\a_A f_A \circ T + \a_B f_B$ (instead of $\a_A f_A + \a_B f_B$). In the current one feature type example, being a cutoff function of $\a_A f_A \circ T + \a_B f_B$ is the same as being a cutoff function of $\a_A f_A + \a_B f_B$ -- both are increasing in $x$. With $N > 1$ feature types, cutoff functions of $\a_A f_A \circ T + \a_B f_B$ and $\a_A f_A + \a_B f_B$ generally do not coincide.

If $d_B$ is a cutoff function of $\a_A f_A \circ T + \a_B f_B$, then there is an $s$ such that $d_B(x) = 1$ if and only if $x \geq s$. Let $d_A$ satisfy $d_A \circ T \equiv d_B$. Then $d_A(T(x)) = 1$ if and only if $x \geq s$, which is equivalent to $d_A(x) = 1$ if and only if $x \geq T(s)$. Recall, the group $B$ likelihood of $s$ is the same as the group $A$ likelihood of $T(s)$. Therefore, $d_A$ applied to group $A$ yields the same likelihood cutoff for acceptance as $d_B$ applied to group $B$.

Implementing the modification of common identity described above requires knowing $T$. In practice, a policymaker is likely to know, at best, only certain things about $T$. For example, since group $A$ has better access to test-prep services, the policymaker may know that $T(x) > x$.

This begs the question: Can $T$ somehow be learned from the objects an intervention can condition on -- the feature distributions of the two groups and the DM's beliefs about quality in the two groups? The answer is yes. Shortly, we will see how to compute the feature densities, $p_{A, q}$ and $p_{B, q}$, from those objects. Given $p_{A, q}$ and $p_{B, q}$, $T$ can be computed by solving
\begin{align}\label{integral}
\int_0^{T(x)} p_{A, q}(x) dx = \int_0^x p_{B, q}(x) dx.
\end{align}
\\
\textit{Formal Analysis of the Feature Bias Model with $N$ Feature Types.}

To formally create the feature bias model, fix a parameterization of the original model -- call it the underlying original model -- and augment it with a differentiable cyclically monotone bijection $T : [0, 1]^N \rightarrow [0, 1]^N$. Cyclic monotonicity means there exists a convex function $\varphi : [0, 1]^N \rightarrow \R$ such that $T \equiv \nabla \varphi$. Cyclic monotonicity generalizes monotonicity to when $N > 1$. The underlying original model's feature densities, $p_q$ and $p_u$, apply only to group $B$ in the feature bias model. So relabel them as $p_{B, q}$ and $p_{B, u}$. Condition \eqref{feature_transform}, relating group $A$'s feature densities, $p_{A, q}$ and $p_{A, u}$, to those of group $B$, generalizes to
\begin{align}\label{jacobian}
p_{B, q}(x) = p_{A, q}(T(x))\vert J(T(x)) \vert \mbox{\ \ \ and\ \ \ } p_{B, u}(x) = p_{A, u}(T(x))\vert J(T(x)) \vert,
\end{align}
where $\vert J(T(x)) \vert$ is the determinant of the Jacobian of $T$ at $x$. The feature bias model is now fully parameterized. When $T$ is the identity function, the feature bias model reduces to the underlying original model.

Let $\pi_g$ be the fraction of qualified individuals in group $g$. In the feature bias model, the feature densities of the two groups and the rational beliefs about quality in the two groups can differ, even if both groups have the same fraction of qualified individuals. For group $g$, the feature density and the rational beliefs about quality are $\mu \vert_g \pi_g$ and $f \vert_g \pi_g$, respectively, where
\begin{align*}
\mu(x) \vert_g \pi_g := \pi_g p_{g, q}(x) + (1 - \pi_g) p_{g, u}(x),\\
f(x) \vert_g \pi_g := \frac{\pi_g p_{g, q}(x)}{\pi_g p_{g, q}(x) + (1 - \pi_g) p_{g, u}(x)}.
\end{align*}
When a group's quality profile is $\ol{c}$, the DM's best-response, given rational beliefs about quality in the group, is to accept a member with features $x$ if and only if the group-specific likelihood of $x$ is at least $\left[ \frac{1 - H(c)}{H(c)}\right] \frac{v_u}{v_q}$, which is the $L(c)$ of the underlying original model.

With likelihood functions being group-specific in the feature bias model, a fixed likelihood-cutoff decision policy $\ul{l}$ is actually a \emph{different} decision policy when applied to different groups. However, the incentive provided by $\ul{l}$ remains group-independent. Indeed, group $B$'s incentive to become qualified facing $\ul{l}$ is
\begin{align*}
	& \int_{\left\{x \big\vert \frac{p_{B, q}(x)}{p_{B, u}(x)} \geq l\right\}} \left( p_{B, q}(x) - p_{B, u}(x) \right) dx\\
=	& \int_{\left\{x \big\vert \frac{p_{A, q}(T(x))}{p_{A, u}(T(x))} \geq l\right\} }\left( p_{A, q}(T(x))\vert J(T(x)) \vert - p_{A, u}(T(x))\vert J(T(x)) \vert \right) dx\\
=	& \int_{\left\{x \big\vert \frac{p_{A, q}(x)}{p_{A, u}(x)} \geq l\right\}} \left( p_{A, q}(x) - p_{A, u}(x) \right) dx,
\end{align*}
which is group $A$'s incentive to become qualified facing $\ul{l}$. In particular, the incentive to become qualified facing $\ul{l}$ equals the $C(l)$ of the underlying original model.

So, unconstrained equilibria still correspond to intersections of the graphs of the underlying original model's $C$ and $L$.
\begin{lemma}
A quality profile $\ol{c^*}$ and a likelihood-cutoff decision policy $\ul{l^*}$ comprise an unconstrained equilibrium in the feature bias model if and only if they comprise an unconstrained equilibrium in the underlying original model.
\end{lemma}

Let $(\ol{c^*}, \ul{l^*})$ be an unconstrained equilibrium. Because $\ul{l^*}$ is a different decision policy when applied to different groups, group $A$'s unconstrained equilibrium $(\ol{c^*}, \ul{l^*})$ is technically not the same as group $B$'s unconstrained equilibrium $(\ol{c^*}, \ul{l^*})$. Nevertheless, when both groups play $(\ol{c^*}, \ul{l^*})$, it is reasonable to think of them as playing the same unconstrained equilibrium: Both groups have the same quality profile; the DM applies the same likelihood cutoff for acceptance to both groups; a group $A$ individual has the same expected payoff as a group $B$ individual with the same cost; and the DM's utilities from the two groups are the same.

Next, let us see how $T$ can be learned from the feature densities of the two groups and the DM's beliefs about quality in the two groups. Once $T$ is learned, we can modify common identity as described in the analysis of the one feature type example.

Let $\pi_g$ be the fraction of qualified individuals in group $g$, so that the group's feature density is $\mu \vert_g \pi_g$, and let $f_g$ be the DM's beliefs about quality in the group. Define the estimator $\hat{p}_{g, q}(\mu \vert_g \pi_g, f_g)$ of $p_{g, q}$, where
\begin{align*}
\hat{p}_{g, q}(\mu \vert_g \pi_g, f_g)(x) := \frac{(\mu(x) \vert_g \pi_g) f_g(x)}{\int_{[0, 1]^N} (\mu(z) \vert_g \pi_g) f_g(z) dz}.
\end{align*}
When $\pi_g > 0$ and $f_g$ is rational, $(\mu(x) \vert_g \pi_g) f_g(x) = (\mu(x) \vert_g \pi_g)(f(x) \vert_g \pi_g) = \pi_g p_{g, q}(x)$, which implies $\hat{p}_{g, q}(\mu \vert_g \pi_g, f_g) \equiv p_{g, q}$.

Once $p_{A, q}$ and $p_{B, q}$ are obtained, $T$ can be learned, in the one feature type example, using Equation \eqref{integral}. With $N > 1$ feature types, this approach will generally not work. Instead, we appeal to optimal transport theory.

\begin{definition}
Let $\nu$, $\z$ be measures on $[0, 1]^N$ and $t$ be a map from $[0, 1]^N$ to itself. The pushforward of $\nu$ under $t$ is the measure $t_{\#} \nu$ on $[0, 1]^N$ satisfying $t_{\#} \nu(A) = \nu(t^{-1}(A))$ for all measurable $A \subset [0, 1]^N$. If $t_{\#} \nu \equiv \z$, then we say $t$ is a transport from $\nu$ to $\z$.
\end{definition}
Given a metric $m$ on $[0, 1]^N$, the \emph{optimal transport problem} from $\nu$ to $\z$ under $m$ is to find a transport $t^*$ from $\nu$ to $\z$ that minimizes expected distance under $m$,
\begin{align*}
\int_{[0, 1]^N} m(x, t^*(x)) d \nu(x) = \inf_{\{t : [0, 1]^N \rightarrow [0, 1]^N \vert t_{\#}\nu \equiv \z\}}\ \int_{[0, 1]^N} m(x, t(x)) d \nu(x).
\end{align*}

\begin{polar}\citep{brenier1991polar}
If $\nu$ has a density, then there is a unique cyclically monotone transport $t^*$ from $\nu$ to $\z$. This $t^*$ is the unique solution to the optimal transport problem from $\nu$ to $\z$ under the squared Euclidean metric.
\end{polar}

Since $T$ is a cyclically monotone transport from $p_{B, q}$ to $p_{A, q}$, the Polar Factorization Theorem implies $T$ is the unique solution to the optimal transport problem from $p_{B, q}$ to $p_{A, q}$ under the squared Euclidean metric. More generally, given the feature densities of the two groups, $(\mu_A, \mu_B)$, and the DM's beliefs about quality in the two groups, $(f_A, f_B)$, define $\hat{T}(\mu_A, \mu_B, f_A, f_B)$ to be the unique solution to the optimal transport problem from $\hat{p}_{B, q}(\mu_B, f_B)$ to $\hat{p}_{A, q}(\mu_A, f_A)$ under the squared Euclidean metric.

We are ready to define the modified common identity intervention.

\begin{definition}
Fix a probability weighting $(\a_A, \a_B)$. The optimal transport common identity intervention allows the set, $k_{OTCI}(\mu_A, \mu_B, f_A, f_B)$, consisting of $(d_A, d_B)$, such that $d_A \circ \hat{T}(\mu_A, \mu_B, f_A, f_B)$ and $d_B$ are the same cutoff function of $\a_A f_A \circ \hat{T}(\mu_A, \mu_B, f_A, f_B) + \a_B f_B$.
\end{definition}

The analysis of the one feature type example implies $f_A \circ \hat{T}(\mu_A, \mu_B, f_A, f_B)$ and $f_B$ are comonotonic with group $B$'s likelihood function when $f_A$ and $f_B$ are rational. The proof of Lemma \ref{obj_1} can be adapted to show

\begin{lemma}\label{obj_1_feature}
In the feature bias model, suppose the fraction of qualified individuals in each group is strictly between zero and one, and the DM's beliefs about quality are rational. Then a decision policy pair $(d_A, d_B)$ is allowed by $k_{OTCI}$ if and only if there is a likelihood-cutoff decision policy $\ul{l}$ such that $d_A \equiv \ul{l}$ when $\ul{l}$ is viewed as a decision policy applied to group $A$, and $d_B \equiv \ul{l}$ when $\ul{l}$ is viewed as a decision policy applied to group $B$.
\end{lemma}

\begin{proof}
See the appendix.
\end{proof}

Just as Lemma \ref{obj_1} implied Theorem \ref{obj_2}, so Lemma \ref{obj_1_feature} implies $k_{OTCI}$-constrained equilibria coincide with scenarios in which both groups play the same unconstrained equilibrium:

\begin{theorem}\label{obj_2_feature}
In the feature bias model, a strategy profile $\{(\ol{c_A^*}, d_A^*), (\ol{c_B^*}, d_B^*)\}$ is a $k_{OTCI}$-constrained equilibrium if and only if there is an unconstrained equilibrium $(\ol{c^*}, \ul{l^*})$ such that $(\ol{c_A^*}, d_A^*) \equiv (\ol{c^*}, \ul{l^*})$ when $(\ol{c^*}, \ul{l^*})$ is viewed as a group $A$ unconstrained equilibrium, and $(\ol{c_B^*}, d_B^*) \equiv (\ol{c^*}, \ul{l^*})$ when $(\ol{c^*}, \ul{l^*})$ is viewed as a group $B$ unconstrained equilibrium.
\end{theorem}
 
\subsection{Group-Specific Cost Distributions}

Sometimes, different groups face substantially different costs to becoming qualified. Consider a judge deciding to grant or deny bail based on features that predict if an individual arrested for a violent crime will recidivate, if released. Biological factors may cause males to find it significantly more costly to not recidivate, if released, than females. In such a setting, imposing common identity can yield scenarios that are highly unfair and inefficient.

To see why, let us now allow the two groups to have different cost distributions while keeping everything else about the original model unchanged. Let $H_g$ denote the CDF of group $g$'s cost distribution. Since the feature densities conditional on being qualified and unqualified are unchanged, the incentive to become qualified, given a decision policy, is unchanged and the same for both groups. However, now that the two groups can have different cost distributions, even when both groups have the same incentive to become qualified, they need not have the same fraction of qualified individuals. For example, suppose both groups have incentive $c$ to become qualified. Then the profile of best-response quality choices is $\ol{c}$ for both groups. But under $\ol{c}$, the fractions of qualified individuals in groups $A$ and $B$ are $H_A(c)$ and $H_B(c)$, respectively. If $H_A(c) > H_B(c)$, then group $A$ is more qualified than group $B$.

Since the incentive to become qualified, given a decision policy, is unchanged, a group's best-response quality profile to likelihood-cutoff decision policy $\ul{l}$ is still $\ol{C(l)}$. However, when a group's quality profile is $\ol{c}$, the DM's best-response decision policy, given rational beliefs about quality, is now group-specific. For group $g$, it is $\ul{L_g(c)}$, where $L_g(c):= \left[ \frac{1 - H_g(c)}{H_g(c)} \right]  \frac{v_u}{v_q}$. Therefore, equilibria between group $g$ and an unconstrained DM correspond to intersections of the graphs of $C$ and $L_g$.

\begin{figure}[t]
  \centering
    \centering
    \scalebox{.8}{%
 \executeiffilenewer{InkscapePics/unconstrained_cost.svg}{InkscapePics/unconstrained_cost.pdf}%
 {inkscape -z -D --file=InkscapePics/unconstrained_cost.svg %
 --export-pdf=InkscapePics/unconstrained_cost.pdf --export-latex}%
 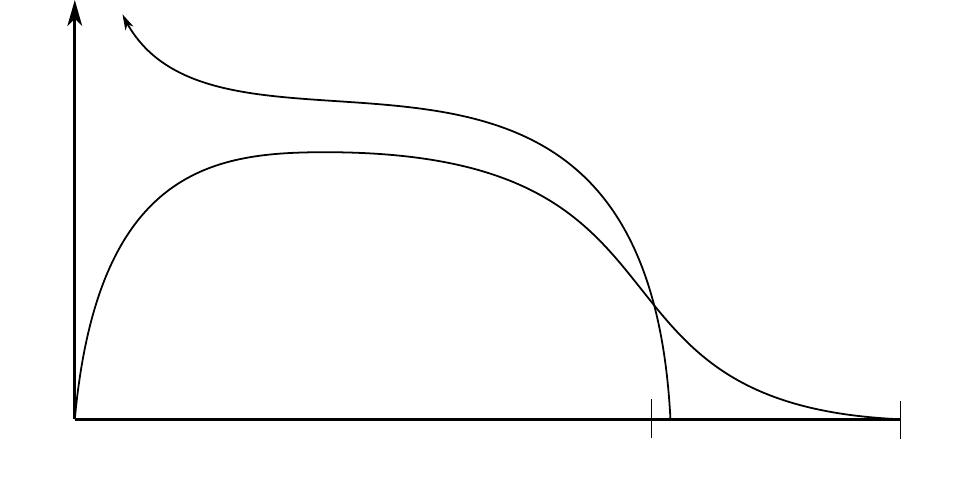%
}
    \caption{Unconstrained equilibria under group-specific cost distributions.}
    \label{fig: unconstrained_cost}
\end{figure}

Figure \ref{fig: unconstrained_cost} depicts a parameterization of the model, in which the cost distributions that group $A$ and group $B$ individuals draw from differ substantially. Notice, $L_B(c_{high})$ is close to $L_B(0)$. If we assume very few group $B$ individuals draw negative costs (i.e., $H_B(0)$ is small), then most group $B$ individuals draw costs exceeding $c_{high}$. By contrast, most group $A$ individuals draw costs less than $c_{low}$. In this example, each group has a unique unconstrained equilibrium. In group $A$'s unconstrained equilibrium, most group $A$ individuals are qualified and, consequently, most group $A$ individuals are accepted via a very low likelihood cutoff $l_A^*$. In group $B$'s unconstrained equilibrium, most group $B$ individuals are unqualified and, consequently, most group $B$ individuals are rejected via a very high likelihood cutoff $l_B^*$.

Next, let us evaluate the impact of common identity. The proof of Lemma \ref{obj_1} remains valid. So, in a common-identity-constrained equilibrium, the DM applies the same likelihood-cutoff decision policy $\ul{l^*}$ to both groups. Both groups then best-respond with the same quality profile $\ol{c^*}$ satisfying $c^* = C(l^*)$. Thus, a common-identity-constrained equilibrium corresponds to a point $(c^*, l^*)$ on the graph of $C$.

\begin{lemma}\label{different_cost_eq}
Let $(c^*, l^*)$ be a point on the graph of $C$ corresponding to a common-identity-constrained equilibrium. Then it must lie between the graphs of $L_A$ and $L_B$ -- that is, $\min\{L_A(c^*), L_B(c^*)\} \leq l^*\leq \max\{L_A(c^*), L_B(c^*)\}$.
\end{lemma}

To prove Lemma \ref{different_cost_eq}, consider an individual with features $x$ satisfying $\frac{p_q(x)}{p_u(x)} > \max\{L_A(c^*), L_B(c^*)\}$. Regardless of which group the individual belongs to, the DM is strictly better off accepting them than rejecting them. This implies $l^*\leq \max\{L_A(c^*), L_B(c^*)\}$. A similar argument implies $l^* \geq \min\{L_A(c^*), L_B(c^*)\}$. 

Now, go back to the example depicted in Figure \ref{fig: unconstrained_cost}. Let $\{(\ol{c^*}, \ul{l^*}), (\ol{c^*}, \ul{l^*})\}$ be a common-identity-constrained equilibrium. For $(c^*, l^*)$ to be between the graphs of $L_A$ and $L_B$, as required by Lemma \ref{different_cost_eq}, it must be that $l^* \in [l_A^*, l_B^*]$, which then implies $c_{low} < C(l^*) < c_{high}$. This means most group $A$ individuals are qualified and most group $B$ individuals are unqualified.

Finally, set $C(1)$ not close to 1. For example, suppose $\int_{\left\{x \vert \frac{p_q(x)}{p_u(x)} \geq 1 \right\}} p_q(x)dx = 0.55$ and $\int_{\left\{x \vert \frac{p_q(x)}{p_u(x)} \geq 1 \right\}} p_u(x)dx = 0.45$, so that $C(1) = 0.1$. If $l^* \geq 1$, then despite most group $A$ individuals being qualified, at least 45\% of qualified group $A$ individuals are rejected. If $l^* \leq 1$, then despite most group $B$ individuals being unqualified, at least 45\% of unqualified group $B$ individuals are accepted. Both scenarios are highly unfair and inefficient.

\subsection{Missing Observations in the Training Dataset}

Machine learning algorithms are trained on datasets consisting of samples of past populations, with each data point $(g, x, y)$ specifying a past individual's group $g$, features $x$, and revealed quality $y \in \{q, u\}$. Even if past populations are similar to the current one, samplings that suffer some form of missing observations can distort the DM's beliefs about quality. Moreover, the distribution of features is not directly observed in many applications. In these cases, the training dataset's empirical feature distribution is often used to proxy for the true feature distribution. How do distortions to the feature distribution and the beliefs about quality, resulting from missing observations, affect common identity, as well as the group fairness interventions of the algorithmic fairness literature?

\subsubsection{Sample Bias and Feature Distribution Shift}\label{shift}

In practice, the probability an individual with data $(g, x, y)$ is sampled often depends on $(g, x, y)$. One common form of dependence is when the sampling probability is independent of $y$ conditional on $(g, x)$. For example, in epidemiology, where $y$ is a health condition and $x$ is a vector of biomarkers, individuals in certain groups are over- or under-sampled, specifically due to their biomarkers. This results in a particular form of sample bias.

To create the sample bias model, fix a parameterization of the original model -- call it the underlying original model -- and augment it with a pair of sampling densities, $m_A$ and $m_B$, one for each group. A sampling density is a full-support density over the feature space. Sample bias is operationalized by incorporating a group's sampling density into the computation of the group's feature density as well as the rational beliefs about quality in the group.

Let $\pi_g$ be the fraction of qualified individuals in group $g$. The sample-biased feature density of group $g$ is $\mu \vert_g \pi_g$, where
  \begin{align*}
  \mu(x)\vert_g \pi_g := \frac{\left[ \pi_g p_q(x) + (1-\pi_g) p_u(x) \right] m_g(x)}{\int_{[0, 1]^N} \left[ \pi_g p_q(z) + (1-\pi_g) p_u(z) \right]m_g(z)dz}.
  \end{align*}
The sample-biased feature densities of the two groups can differ, even if both groups have the same fraction of qualified individuals. The sample-biased feature density of a group can also differ from the group's true feature density. However, the rational beliefs about quality in a group are unaffected by incorporating the group's sampling density,
  \begin{align*}
  \frac{\pi_g p_q(x)m_g(x)}{\left[\pi_g p_q(x) + (1-\pi_g) p_u(x)\right] m_g(x)}
   													 = \frac{\pi_g p_q(x)}{\pi_g p_q(x) + (1-\pi_g) p_u(x)}.
  \end{align*}
When the feature density is distorted, but the beliefs about quality are not, it is known as \emph{feature distribution shift} or \emph{covariate shift}.

In the sample bias model, we modify the definitions of unconstrained equilibrium, intervention, and intervention-constrained equilibrium by replacing the true feature densities with the sample-biased ones. Since the rational beliefs about quality are unchanged, 

\begin{lemma}\label{sample_original}
A quality profile $\ol{c^*}$ and a likelihood-cutoff decision policy $\ul{l^*}$ comprise an unconstrained equilibrium in the sample bias model if and only if they comprise an unconstrained equilibrium in the underlying original model.
\end{lemma}

Lemma \ref{obj_1} is still true. This is because the set of decision policy pairs allowed by common identity does not depend on the feature densities of the two groups. Thus, replacing the true feature densities with the sample-biased ones has no effect. We may now conclude that in the sample bias model, common-identity-constrained equilibria continue to coincide with scenarios in which both groups play the same unconstrained equilibrium:

\begin{theorem}\label{sample_CI}
In the sample bias model, a strategy profile $\{(\ol{c_A^*}, d_A^*), (\ol{c_B^*}, d_B^*)\}$ is a common-identity-constrained equilibrium if and only if there is an unconstrained equilibrium $(\ol{c^*}, \ul{l^*})$ such that $(\ol{c_A^*}, d_A^*) \equiv (\ol{c_B^*}, d_B^*) \equiv (\ol{c^*}, \ul{l^*})$.
\end{theorem}

By contrast, the group fairness interventions of the algorithmic fairness literature are highly sensitive to the feature densities of the two groups. For example,

\begin{definition}
The equality of opportunity intervention requires the probability of acceptance, conditional on being qualified, to be equal across groups: $k_{EO}(\mu_A, \mu_B, f_A, f_B) =$
\begin{align*}
\left\{ (d_A, d_B)\ \bigg\vert\ \frac{\int_{[0, 1]^N} d_A(x) \mu_A(x) f_A(x)dx}{\int_{[0, 1]^N} \mu_A(x) f_A(x)dx} = \frac{\int_{[0, 1]^N} d_B(x) \mu_B(x) f_B(x)dx}{\int_{[0, 1]^N} \mu_B(x) f_B(x)dx} \right\}.
\end{align*}
%If $f_g \equiv 0$, then set $\frac{\int_{[0, 1]^N} d_g(x) \mu_g(x) f_g(x)dx}{\int_{[0, 1]^N} \mu_g(x) f_g(x)dx}$ equal to zero.
\end{definition}

This means even if a group fairness intervention performs as well as common identity in the original model, it need not in the sample bias model.

\begin{proposition}\label{equal_opp}
In the sample bias model, when $m_A$ and $m_B$ are uniform, a strategy profile $\{(\ol{c_A^*}, d_A^*), (\ol{c_B^*}, d_B^*)\}$ is a $k_{EO}$-constrained equilibrium if and only if there is an unconstrained equilibrium $(\ol{c^*}, \ul{l^*})$ such that $(\ol{c_A^*}, d_A^*) \equiv (\ol{c_B^*}, d_B^*) \equiv (\ol{c^*}, \ul{l^*})$. However, in general, there exist parameterizations of the sample bias model that have $k_{EO}$-constrained equilibria involving unequally qualified groups.
\end{proposition}

\begin{proof}
See the appendix for a proof of the first part.
\end{proof}

When the sampling densities are uniform, the sample-biased feature densities of the two groups are the true feature densities. The sample bias model then reduces to the underlying original model.

In general, the sample-biased feature densities of the two groups can differ from the true feature densities in arbitrary ways: Let $\pi_g$ be the fraction of qualified individuals in group $g$, and let $\mu$ be an arbitrary full-support feature density. Then $\mu \vert_g \pi_g \equiv \mu$, if group $g$'s sampling density is $m_g$, where
\begin{align*}
m_g(x) := \frac{\frac{\mu(x)}{\pi_g p_q(x) + (1-\pi_g) p_u(x)}}{\int_{[0, 1]^N} \frac{\mu(z)}{\pi_g p_q(z) + (1-\pi_g) p_u(z)} dz}.
\end{align*}

To prove the second part of Proposition \ref{equal_opp}, fix an underlying original model with distinct unconstrained equilibria, $(\ol{c_A^*}, \ul{l_A^*})$ and $(\ol{c_B^*}, \ul{l_B^*})$. Define $\tau_A \in (0, 1)$ as a group's probability of acceptance, conditional on being qualified, in the underlying original model, when it plays $(\ol{c_A^*}, \ul{l_A^*})$ with the DM. Then
\begin{align*}
\tau_A = \frac{\int_{[0, 1]^N} \ul{l_A^*}(x) (\mu(x) \vert H(c_A^*)) (f(x) \vert H(c_A^*))dx}{\int_{[0, 1]^N} (\mu(x) \vert H(c_A^*)) (f(x) \vert H(c_A^*))dx}.
\end{align*}
Since $\ul{l_B^*}(x) = 1$ if and only if the likelihood of $x$ is at least $l_B^*$, and $l_B^* \in (0, \infty)$, if a full-support feature density $\mu$ places sufficiently large weights on $x$ with likelihoods above (below) $l_B^*$, then
\begin{align*}
\tau_{\mu} := \frac{\int_{[0, 1]^N} \ul{l_B^*}(x) \mu(x) (f(x) \vert H(c_B^*))dx}{\int_{[0, 1]^N} \mu(x) (f(x) \vert H(c_B^*))dx} > (<)\ \tau_A.
\end{align*}
By taking an appropriate weighted average of a $\mu_{high}$ satisfying $\tau_{\mu_{high}} > \tau_A$ and a $\mu_{low}$ satisfying $\tau_{\mu_{low}} < \tau_A$, we can find a full-support feature density $\mu^*$ satisfying $\tau_{\mu^*} = \tau_A$. Fix such a $\mu^*$. By the earlier analysis, there exists a group $B$ sampling density $m_B$ under which $\mu\vert_B H(c_B^*) \equiv \mu^*$.

Augment the underlying original model with the uniform sampling density for group $A$ and sampling density $m_B$ for group $B$. By Lemma \ref{sample_original}, $(\ol{c_A^*}, \ul{l_A^*})$ and $(\ol{c_B^*}, \ul{l_B^*})$ are unconstrained equilibria in the sample bias model. By design, if each group $g$ plays $(\ol{c_g^*}, \ul{l_g^*})$ with the DM, then group $A$'s ($B$'s) probability of acceptance, conditional on being qualified, in the sample bias model, is $\tau_A$ ($\tau_{\mu^*}$). Since $\tau_{\mu^*} = \tau_A$, $\{(\ol{c_A^*}, \ul{l_A^*}), (\ol{c_B^*}, \ul{l_B^*})\}$ is a $k_{EO}$-constrained equilibrium in the sample bias model -- one that involves unequally qualified groups because $H(c_A^*) \neq H(c_B^*)$.

\subsubsection{Label Bias}

Recall, training datasets consist of samples of past individuals whose previously unobserved qualities have been revealed. In a setting like lending, where qualified denotes will not default, generating such samples is straightforward. Eventually, every past borrower is observed to have defaulted or not. In other settings, getting data about quality is not so straightforward.

Consider a firm that wants to develop a hiring algorithm to target applicants who are qualified for a complex job. In such a setting, being qualified can be hard to observe even ex-post. As a result, the firm may choose to use its own past hiring decisions for similar jobs, which are easy to observe, to proxy for quality. In this case, the training dataset consists of data points, $(g, x, \hat{y})$, each specifying a past applicant's group $g$, features $x$, and hiring decision $\hat{y}$, where hired and not hired are proxies for qualified and unqualified, respectively. The resulting machine-learning-generated beliefs about the hiring decision may not be close to correct, if treated as beliefs about quality. To make matters worse, the joint distribution of hiring decision and quality often depends on group membership. For example, in certain industries, compared to men, women were less likely to be hired in the past, even after controlling for quality, due to past employer biases. In this case, beliefs about quality, trained on data about past hiring decisions, may lead to too few women being hired.

When the DM's beliefs about quality are incorrect because the training dataset uses a proxy for quality, it is called \emph{label bias}. To create the label bias model, fix a parameterization of the original model -- call it the underlying original model -- and augment it with proxy-qualities, $\hat{q}$ and $\hat{u}$. Let $y \in \{q, u\}$ and $\hat{y} \in \{\hat{q}, \hat{u}\}$ denote an individual's quality and proxy-quality, respectively.

To motivate how to model the joint distribution of $(y, \hat{y})$, let us revisit the hiring example. Holding group $g$ and quality $y$ fixed, it is plausible that a past employer -- even one biased against certain groups -- was more likely to hire someone if the likelihood of their features $x$ was higher. Additionally, the past employer's hiring decisions may have been influenced by signals of quality not captured in the applicant's features $x$. It is plausible, holding group $g$ and features $x$ fixed, that a qualified applicant was more likely to generate stronger signal values and be hired than an unqualified applicant. Therefore, we make the following assumption:

\begin{assumption}\label{bias_assumption}
For each group $g$, there exist strictly increasing functions $\p_{g, q}, \p_{g, u} : [0, \infty] \rightarrow [0, 1]$ such that, for all features $x \in [0, 1]^N$,
\begin{align*}
\Pr(\hat{y} = \hat{q} \vert g, x, y = q) = \p_{g, q}\left( \frac{p_q(x)}{p_u(x)}\right) \geq \p_{g, u}\left( \frac{p_q(x)}{p_u(x)}\right) = \Pr(\hat{y} = \hat{q} \vert g, x, y = u).
\end{align*}
\end{assumption}

In the label bias model, we modify the definitions of unconstrained equilibrium, intervention, and intervention-constrained equilibrium by replacing beliefs about quality with beliefs about proxy-quality.

Let $\pi_g$ be the fraction of qualified individuals in group $g$. Assumption \ref{bias_assumption} implies the DM's rational beliefs about proxy-quality in group $g$ are $\hat{f} \vert_g \pi_g$, where
\begin{align}\label{bias_belief}
\hat{f}(x) \vert_g \pi_g := \frac{\pi_g p_q(x) \p_{g, q}\left( \frac{p_q(x)}{p_u(x)}\right) + (1 - \pi_g) p_u(x) \p_{g, u}\left( \frac{p_q(x)}{p_u(x)}\right)}{\pi_g p_q(x) + (1 - \pi_g) p_u(x)}.
\end{align}
Equation \eqref{bias_belief} implies the DM's rational beliefs about proxy-quality in the two groups can differ, even if both groups have the same fraction of qualified individuals. As a result, each group may now have a different set of unconstrained equilibria, in which case it is impossible to get both groups to play the same unconstrained equilibrium.

However, common identity still induces the DM to provide both groups the same incentive to become qualified. Equation \eqref{bias_belief} implies that $\hat{f}(x) \vert_g \pi_g$ can be rewritten as
\begin{align*}
\frac{\pi_g \frac{p_q(x)}{p_u(x)} \p_{g, q}\left( \frac{p_q(x)}{p_u(x)}\right) + (1 - \pi_g)\p_{g, u}\left( \frac{p_q(x)}{p_u(x)}\right)}{\pi_g \frac{p_q(x)}{p_u(x)} + (1 - \pi_g)}.
\end{align*}
Let $x$ and $x'$ have likelihoods $l$ and $l'$, respectively, with $l < l'$. Then
\begin{align*}
\hat{f}(x') \vert_g \pi_g - \hat{f}(x) \vert_g \pi_g =	&\frac{(\pi_g l' \p_{g, q}(l') + (1 - \pi_g) \p_{g, u}(l'))(\pi_g l + (1 - \pi_g))}{(\pi_g l' + (1 - \pi_g))(\pi_g l + (1 - \pi_g))}\\
	& - \frac{(\pi_g l \p_{g, q}(l) + (1 - \pi_g) \p_{g, u}(l))(\pi_g l' + (1 - \pi_g))}{(\pi_g l' + (1 - \pi_g))(\pi_g l + (1 - \pi_g))}\\
=	&\frac{\pi_g^2l'l(\p_{g, q}(l') - \p_{g, q}(l)) + (1 - \pi_g)^2(\p_{g, u}(l') - \p_{g, u}(l))}{(\pi_g l' + (1 - \pi_g))(\pi_g l + (1 - \pi_g))}\\
	& + \frac{\pi_g(1 - \pi_g)(l'(\p_{g, q}(l') - \p_{g, u}(l)) - l(\p_{g, q}(l) - \p_{g, u}(l')))}{(\pi_g l' + (1 - \pi_g))(\pi_g l + (1 - \pi_g))}.
\end{align*}
In the last expression above, the first term is positive because $\p_{g, q}$ and $\p_{g, u}$ are strictly increasing. Assumption \ref{bias_assumption} ensures that $\p_{g, q}(l') - \p_{g, u}(l) > \max\{\p_{g, q}(l) - \p_{g, u}(l'), 0\}$. Since $l' > l \geq 0$, the second term is nonnegative. Thus, $\hat{f} \vert_g \pi_g$ is comonotonic with the likelihood function. Using the same reasoning for Lemma \ref{obj_1}, we have
\begin{lemma}\label{obj_1_label}
In the label bias model, suppose the DM's beliefs about proxy-quality are rational. Then a decision policy pair is allowed by common identity if and only if both components are the same likelihood-cutoff decision policy.
\end{lemma}

Let $\{(\ol{c_A^*}, d_A^*), (\ol{c_B^*}, d^*_B)\}$ be a common-identity-constrained equilibrium. Lemma \ref{obj_1_label} implies there is an $\ul{l^*}$ such that $d_A^* \equiv d_B^* \equiv \ul{l^*}$. Therefore, both groups have the same incentive $C(l^*)$ to become qualified. So both groups are equally qualified, $c_A^* = c_B^* = C(l^*)$. Due to label bias, the DM may still have worse beliefs about proxy-quality in, say, group $B$ -- i.e., $\hat{f}(x)\vert_B H(c_B^*) < \hat{f}(x)\vert_A H(c_A^*)$ for all features $x$. If so, common identity prevents a group $B$ individual from being worse off than a group $A$ individual with the same cost, which would happen if the DM were unconstrained.

\section{Some Concluding Remarks}

\subsection{Statistical Discrimination and Group Fairness}\label{group_fair}

Consider a situation in which it is observed that the DM's machine-learning-generated beliefs about quality are worse for group $B$ than for group $A$. Consequently, the DM is observed to apply a tougher decision policy to group $B$ than to group $A$. In this paper, we viewed such a situation through the lens of a statistical discrimination model. The observations were interpreted as symptoms of group $B$ playing a worse unconstrained equilibrium than group $A$. A model-driven policy goal was then articulated to guide the search for an intervention: Seek an intervention with the property that intervention-constrained equilibria coincide with scenarios in which both groups play the same unconstrained equilibrium.

By contrast, the algorithmic fairness literature does not focus on first providing an economic model of the situation. Instead, what often happens is a group fairness concept is axiomatically chosen, which then determines the choice of intervention. For example, one could find it unethical for different groups to have different acceptance rates. This then motivates imposing an intervention that satisfies the group fairness concept of statistical parity.

\begin{definition}
The statistical parity intervention requires the probability of acceptance to be equal across groups: $k_{SP}(\mu_A, \mu_B, f_A, f_B) = $
\begin{align*}
\left\{ (d_A, d_B)\ \bigg\vert\ \int_{[0, 1]^N} d_A(x) \mu_A(x) dx = \int_{[0, 1]^N} d_B(x) \mu_B(x) dx \right\}.
\end{align*}
\end{definition}

Compared to my approach to addressing algorithmic bias, there are obvious advantages to the group fairness approach. Group fairness concepts ``can be readily applied across contexts, with little domain-specific knowledge" \citep{corbett2023measure}. Moreover, group fairness concepts are defined using the feature densities of the two groups, $(\mu_A, \mu_B)$, and the DM's beliefs about quality in the two groups, $(f_A, f_B)$. Since interventions can fully condition on those objects, it is trivial to design an intervention to satisfy a group fairness concept -- just use the definition of the group fairness concept as the constraint on what kinds of decision policy pairs are allowed. This is exactly how the statistical parity intervention above and the equality of opportunity intervention in Section \ref{shift} are designed.

However, recent research has highlighted drawbacks of the group fairness approach to algorithmic bias. \citet{kleinberg2016inherent} and \citet{chouldechova2017fair} show that some group fairness concepts are mutually incompatible. Others have shown that group fairness interventions can lead to Pareto-inferior outcomes and greater unfairness when measured in an application-appropriate way \citep{corbett2017algorithmic, liu2018delayed, hu2020fair}. Instead, \citet{corbett2023measure} ``advocate for a consequentialist perspective that directly grapples with the difficult policy trade-offs inherent to many algorithmically guided decisions" and warn that
\begin{quote}
Formal fairness criteria are thus often at odds with policy goals and, perversely, can harm the very same groups one ostensibly sought to protect by developing and adopting axiomatic notions of fairness.
\end{quote}

As a demonstration, suppose the situation described in the beginning of this section is, in reality, explained by our model of statistical discrimination. A policymaker, instead of bothering with the model, takes the group fairness approach to addressing the situation. The policymaker finds it unethical that group $B$ has a lower acceptance rate than group $A$ and imposes the statistical parity intervention. Does this intervention achieve the policy goal motivated by the model that actually explains the situation? That is, do statistical-parity-constrained equilibria coincide with scenarios in which both groups play the same unconstrained equilibrium?

The answer is no. Proposition 4 in \citet{coate1993will} implies that there exist statistical-parity-constrained equilibria that involve unequally qualified groups. In fact, \citet{coate1993will} find that, when the DM statistically discriminates against group $B$, imposing the statistical parity intervention can perversely cause group $B$, under best-response dynamics, to become even less qualified.

%\begin{comment}

Or, consider the group fairness intervention designed by \citet{strack2024privacy}: Given the feature densities of the two groups, $(\mu_A, \mu_B)$, and the DM's beliefs about quality in the two groups, $(f_A, f_B)$, create a pair of quantile signals $q_A(\mu_A, f_A) : [0, 1]^N \rightarrow [0, 1]$ and $q_B(\mu_B, f_B) : [0, 1]^N \rightarrow [0, 1]$. For each group $g$, $q_g(\mu_g, f_g)(x)$ is the quantile of a member with features $x$, when ranked according to $f_g$. The quantile signal intervention allows $(d_A, d_B)$ if and only if $d_A$ is the same cutoff function of $q_A(\mu_A, f_A)$ as $d_B$ is of $q_B(\mu_B, f_B)$.

The quantile signal intervention admits an information-design interpretation in which the DM only observes the individual's quantile signal value. Therefore, the decision to accept an individual can depend on that individual's group and features only up to the quantile signal value they induce.

The motivation for the quantile signal intervention is to satisfy statistical parity while preserving privacy about group membership. When the DM only observes an individual's quantile signal value, the posterior belief about that individual's group remains at the prior, $(\l_A, \l_B)$, preserving privacy about group membership. Moreover, any decision policy that depends only on the quantile signal value equalizes the probability of acceptance across groups, satisfying statistical parity.

Since the quantile signal intervention satisfies statistical parity, any decision policy pair allowed by it is also allowed by the statistical parity intervention. Conversely, in a statistical-parity-constrained equilibrium, the decision policy pair chosen by the DM is allowed by the quantile signal intervention. This implies that the set of quantile-signal-constrained equilibria is exactly the set of statistical-parity-constrained equilibria. Proposition 4 in \citet{coate1993will} now implies there exist quantile-signal-constrained equilibria that involve unequally qualified groups.

Recall, the DM's utility from applying the decision policy pair $(d_A, d_B)$ to the two groups is $\l_A V(d_A, \mu_A, f_A) + \l_B V(d_B, \mu_B, f_B)$. \citet{liang2024algorithm} introduces a policymaker who has utility
\begin{align*}
\l_A V(d_A, \mu_A, f_A) + \l_B V(d_B, \mu_B, f_B) - \g_{\D} \vert V(d_A, \mu_A, f_A)  - V(d_B, \mu_B, f_B) \vert.
\end{align*}
Here, $\g_{\D} \geq 0$ measures the policymaker's concern for group fairness. It is inspired by the myriad group fairness concepts that involve equalizing, across groups, some integral of the distribution of decisions. For statistical parity and equality of opportunity, the integrals are
\begin{align*}
\int_{[0, 1]^N} d_g(x) \mu_g(x) dx & \mbox{\ \ \ \ \ \ \ and\ \ \ \ \ \ \ } \frac{\int_{[0, 1]^N} d_g(x) \mu_g(x) f_g(x) dx}{\int_{[0, 1]^N} \mu_g(x) f_g(x) dx},
\end{align*}
respectively. Whereas in \citet{liang2024algorithm} it is
\begin{align*}
V(d_g, \mu_g, f_g) = \int_{[0, 1]^N} d_g(x) \mu_g(x) \left[ f_g(x)v_q - \left(1- f_g(x)\right) v_u\right] dx.
\end{align*}

\begin{figure}[t]
  \centering
    \centering
    \scalebox{.8}{%
 \executeiffilenewer{InkscapePics/stat_disc_eq.svg}{InkscapePics/stat_disc_eq.pdf}%
 {inkscape -z -D --file=InkscapePics/stat_disc_eq.svg %
 --export-pdf=InkscapePics/stat_disc_eq.pdf --export-latex}%
 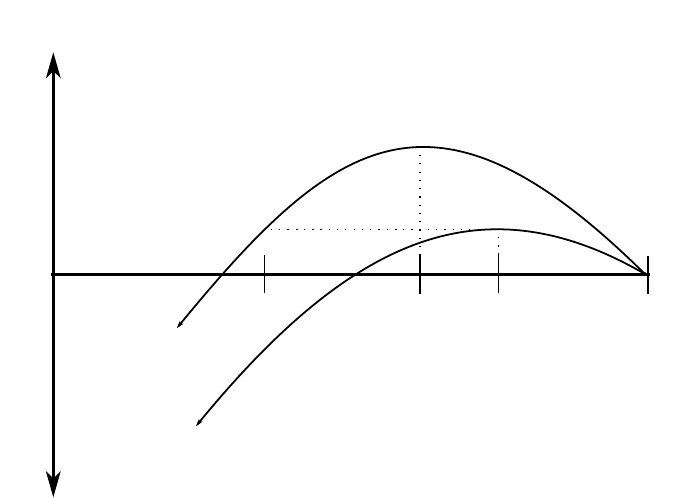%
}
    \caption{Depiction of $V(\ul{l}, \mu \vert H(c_A^*), f \vert H(c_A^*))$ and $V(\ul{l}, \mu \vert H(c_B^*), f \vert H(c_B^*))$ as functions of $l$. Since $(\ol{c_A^*}, \ul{l_A^*})$ Pareto-dominates $(\ol{c_B^*}, \ul{l_B^*})$, we have $l_A^* < l_B^*$, $V(\ul{l_A^*}, \mu \vert H(c_A^*), f \vert H(c_A^*)) > V(\ul{l_B^*}, \mu \vert H(c_B^*), f \vert H(c_B^*))$, and $V(\ul{l}, \mu \vert H(c_A^*), f \vert H(c_A^*)) > V(\ul{l}, \mu \vert H(c_B^*), f \vert H(c_B^*))$ for $l < \infty$.}
    \label{fig: unfair_eq}
\end{figure}

Suppose group $A$'s unconstrained equilibrium, $(\ol{c_A^*}, \ul{l_A^*})$, Pareto-dominates group $B$'s unconstrained equilibrium, $(\ol{c_B^*}, \ul{l_B^*})$. If the policymaker has no concern for group fairness, $\g_{\D} = 0$, then the decision policy pair chosen by an unconstrained DM, $(\ul{l_A^*}, \ul{l_B^*})$, also maximizes the policymaker's utility. However, as $\g_{\D}$ increases, the policymaker does not respond by forcing the DM to make it easier for group $B$ individuals to be accepted. Instead, the policymaker's utility is maximized by decreasing or increasing group $A$'s likelihood cutoff for acceptance. In particular, for all $\g_{\D}$ sufficiently large, the policymaker's utility can be maximized by applying the low likelihood cutoff $l_A^{min}$ to group $A$, satisfying $V(\ul{l_A^{min}}, \mu \vert H(c_A^*), f \vert H(c_A^*)) = V(\ul{l_B^*}, \mu \vert H(c_B^*), f \vert H(c_B^*))$. See Figure \ref{fig: unfair_eq}. 

Thus, despite statistical discrimination against group $B$, it is possible the more concerned the policymaker is about group fairness, the easier it is for group $A$ individuals to be accepted. Meanwhile, the decision policy for group $B$ remains unchanged. 

%In an online appendix, \citet{liang2024algorithm} allow the policymaker's concern for group fairness to relate to something different from $V$. As the authors suggest, this paves the way for $\vert V(\mu \vert \pi_A, f_A, d_A)  - V(\mu \vert \pi_B, f_B, d_B) \vert$ to be replaced with the acceptance rate gap. But as we have seen, it may also not be appropriate for a policymaker, concerned about statistical discrimination, to focus on equalizing acceptance rates.

%To be clear, there is nothing intrinsically wrong with trying to achieve a group fairness goal. In fact, Theorem \ref{equal_opp} states that, at least in the original model, a temporary program of the equality of opportunity intervention ensures both groups play the same equilibrium with the DM. The issue is that group fairness goals are often implemented without any economic model of the individuals being acted upon. This is ironic, given that it is for their benefit that these concepts were developed in the first place.
%\end{comment}

\subsection{Common Identity and the Law}

In this paper, we studied interventions against statistical discrimination by a DM with verifiable beliefs, such as those generated by machine learning, rather than by humans. We analyzed a belief-contingent intervention, common identity, and showed that it can be effective at combating machine-assisted statistical discrimination. Moreover, the performance of common identity is robust to the training dataset exhibiting various statistical biases, including feature bias, sample bias, and label bias.

Implementing common identity requires the right to audit a machine learning algorithm's training data, training procedure, and output. Around the world, laws governing artificial intelligence (AI) have been passed, mandating algorithmic audits to ensure compliance with various regulations. For now, none of those regulations are common identity. However, many of the laws are motivated by algorithmic bias.

In 2021, New York City passed Local Law 144, regulating the use of automated employment decision tools by employers and employment agencies. Such tools are subject to annual independent bias audits. One of the key metrics is the scoring rate: For each group $g$, the scoring rate is the fraction of group $g$ individuals with an $f_g(x)$ exceeding the median or some desirability threshold \citep{wright2024null}.

The Colorado Artificial Intelligence Act was a comprehensive law protecting Colorado residents from algorithmic discrimination in employment, education, housing, finance, healthcare, and essential government services. Developers of high-risk AI systems were required to implement various data governance measures and to provide information necessary for deployers to perform impact assessments. It was repealed and replaced with the Automated Decision-Making Technology Act, which will take effect on January 1, 2027. While the new law is substantially narrower in scope, one important addition is that any consumer, who receives an adverse decision, has the right to a meaningful human review and reconsideration of the consequential decision.

Legislation at the federal level has stalled. The proposed Algorithm Accountability Act of 2023 mandates that companies conduct impact assessments for automated decision systems and augmented critical decision processes. However, the act did not receive a vote after previous versions failed to pass in 2019 and 2022. In fact, the One Big Beautiful Bill Act (H.R.1) originally contained a provision for a 10-year federal moratorium on state and local regulations of AI. That provision was removed by the Senate on July 1, 2025.

The EU Artificial Intelligence Act, adopted in 2024, is a comprehensive legal framework for regulating AI systems. In view of possible algorithmic biases, training, validation, and testing datasets are subject to data governance and management practices. High-risk AI systems that continue to learn after being deployed must eliminate or reduce, as far as possible, the risk of possibly biased outputs. High-risk AI systems must also be designed to include appropriate human-machine interface tools, so they can be effectively overseen by natural persons.

These laws suggest that at least in some jurisdictions, the information that developers and deployers of high-risk AI systems must preserve and make accessible is sufficient to implement common identity.

\section{Appendix}

\begin{proof}[Proof of Proposition \ref{group_blind}]

\begin{lemma}\label{obj_1_blind}
In the separable model, when $G$ is empty, suppose the fraction of qualified individuals in each group is strictly between zero and one, and the DM's beliefs about quality are rational. Then the DM's best-response under group-blinding is to apply the same likelihood-cutoff decision policy to both groups.
\end{lemma}

Just as Lemma \ref{obj_1} implied Theorem \ref{obj_2}, so Lemma \ref{obj_1_blind} implies Proposition \ref{group_blind}. So, to prove Proposition \ref{group_blind}, it suffices to prove Lemma \ref{obj_1_blind}.

To prove Lemma \ref{obj_1_blind}, it suffices to prove that, given rational beliefs about quality, if the DM is weakly better off accepting all individuals with features $x$, then the DM is strictly better off accepting all individuals with features that have strictly higher likelihood than $x$.

Since $G$ is empty, the feature densities conditional on quality are not group-specific, and it is well-defined to write $p_q(x)$, $p_u(x)$, and $\mu \vert \pi_g$.

Let $x$ and $x'$ satisfy $\frac{p_q(x')}{p_u(x')} > \frac{p_q(x)}{p_u(x)}$, and let $\pi_g \in (0, 1)$ be the fraction of qualified individuals in group $g$. Given rational beliefs about quality, the condition that the DM is weakly better off accepting all individuals with features $x$ is
\begin{align*}
			&\sum_{g \in \{A, B\}} \l_g (\mu(x) \vert \pi_g) \left[ \left(f(x)\vert \pi_g\right)v_q - \left(1- f(x)\vert \pi_g\right) v_u\right] \geq 0\\
\Leftrightarrow	&\sum_{g \in \{A, B\}} \l_g \left[ \pi_g p_q(x) v_q - (1 - \pi_g) p_u(x) v_u\right] \geq 0\\
\Leftrightarrow	& \frac{p_q(x)}{p_u(x)} \left[ \l_A \pi_A + \l_B \pi_B \right] v_q \geq \left[ \l_A (1 - \pi_A) + \l_B (1 - \pi_B) \right] v_u.
\end{align*}
Since $\frac{p_q(x')}{p_u(x')} > \frac{p_q(x)}{p_u(x)}$, the DM is strictly better off accepting all individuals with features $x'$.
\end{proof}

\begin{proof}[Proof of Lemma \ref{obj_1_feature}]
Let $\pi_g \in (0, 1)$ be the fraction of qualified individuals in group $g$. Then $\hat{p}_{g, q}(\mu \vert_g \pi_g, f \vert_g \pi_g) \equiv p_{g, q}$ and $\hat{T}(\mu \vert_A \pi_A, \mu \vert_B \pi_B$, $f \vert_A \pi_A, f \vert_B \pi_B) \equiv T$.

Let $(d_A, d_B) \in k_{OTCI}(\mu \vert_A \pi_A, \mu \vert_B \pi_B, f \vert_A \pi_A, f \vert_B \pi_B)$. Then there exists an $s \in [0, 1]$ such that $d_B(x) = 1$ if and only if $\a_A f(T(x)) \vert_A \pi_A + \a_B f(x) \vert_B \pi_B \geq s$, and $d_A(T(x)) = 1$ if and only if $\a_A f(T(x)) \vert_A \pi_A + \a_B f(x) \vert_B \pi_B \geq s$. Since $\frac{p_{B, q}(x)}{p_{B, u}(x)} = \frac{p_{A, q}(T(x))}{p_{A, u}(T(x))}$ and $f(x) \vert_g \pi_g$ is comonotonic with $\frac{p_{g, q}(x)}{p_{g, u}(x)}$ for both groups $g$, we have $\a_A f_A(T(x)) \vert_A \pi_A + \a_B f_B(x) \vert_B \pi_B$ is comonotonic with $\frac{p_{B, q}(x)}{p_{B, u}(x)}$. Thus, there exists an $l \in [0, \infty]$, such that $d_B(x) = 1$ if and only if $\frac{p_{B, q}(x)}{p_{B, u}(x)} \geq l$ and $d_A(T(x)) = 1$ if and only if $\frac{p_{B, q}(x)}{p_{B, u}(x)} \geq l$. Since $\frac{p_{B, q}(x)}{p_{B, u}(x)} = \frac{p_{A, q}(T(x))}{p_{A, u}(T(x))}$, therefore $d_A(T(x)) = 1$ if and only if $\frac{p_{A, q}(T(x))}{p_{A, u}(T(x))} \geq l$. Since $T$ is a bijection, $d_A(T(x)) = 1$ if and only if $\frac{p_{A, q}(T(x))}{p_{A, u}(T(x))} \geq l$ is equivalent to $d_A(x) = 1$ if and only if $\frac{p_{A, q}(x)}{p_{A, u}(x)} \geq l$. This proves $d_g$ is $\ul{l}$ applied to group $g$, for both groups $g$.

The argument can be reversed. If there is an $\ul{l}$ such that $d_g$ is $\ul{l}$ applied to group $g$, for both groups $g$, then $(d_A, d_B) \in k_{OTCI}(\mu \vert_A \pi_A, \mu \vert_B \pi_B, f \vert_A \pi_A, f \vert_B \pi_B)$.
\end{proof}

\begin{proof}[Proof of the first part of Proposition \ref{equal_opp}]
 
\begin{lemma}\label{obj_1_opp}
In the sample bias model, when $m_A$ and $m_B$ uniform, suppose the fraction of qualified individuals in each group is strictly between zero and one, and the DM's beliefs about quality are rational. Then $k_{EO}$ allows any decision policy pair in which both components are the same likelihood-cutoff decision policy. Moreover, the DM's best-response under $k_{EO}$ is such a decision policy pair.
\end{lemma}

Just as Lemma \ref{obj_1} implied Theorem \ref{obj_2}, so Lemma \ref{obj_1_opp} implies the first part of Proposition \ref{equal_opp}. So, to prove the first part of Proposition \ref{equal_opp}, it suffices to prove Lemma \ref{obj_1_opp}.

Let $\pi_g \in (0, 1)$ be the fraction of qualified individuals in group $g$. Since $m_g$ is uniform, group $g$'s sample-biased feature density is its true feature density, and it is well-defined to write $\mu \vert \pi_g$. Given $l \in [0, \infty]$, group $g$'s probability of acceptance, conditional on being qualified, under $\ul{l}$ is
\begin{align}\label{true_positive}
\frac{\int_{[0, 1]^N} \ul{l}(x) (\mu(x) \vert \pi_g) (f(x) \vert \pi_g)dx}{\int_{[0, 1]^N} (\mu(x) \vert \pi_g) (f(x) \vert \pi_g)dx}  = \int_{\left\{x \big\vert \frac{p_q(x)}{p_u(x)} \geq l\right\}} p_q(x) dx.
\end{align}
Since the expression on the right does not depend on group membership, the first part of the lemma is proved.

Next, let $(d_A^*, d_B^*)$ be the best-response of the DM under $k_{EO}$, given rational beliefs about quality. Let $r^*$ denote the probability of acceptance, conditional on being qualified, under both $d_A^*$ and $d_B^*$. The best-response condition implies that, for each group $g$,
\begin{align*}
d_g^* 	& \in \argmax_{\mbox{decision policies $d$} \ s.t.\ \frac{\int_{[0, 1]^N} d(x) (\mu(x) \vert \pi_g) (f(x) \vert \pi_g)dx}{\int_{[0, 1]^N} (\mu(x) \vert \pi_g) (f(x) \vert \pi_g) dx} = r^*} V(d, \mu \vert \pi_g, f \vert \pi_g)\\
		& = \argmax_{\mbox{decision policies $d$} \ s.t.\ \frac{\int_{[0, 1]^N} d(x) (\mu(x) \vert \pi_g) (f(x) \vert \pi_g)dx}{\int_{[0, 1]^N} (\mu(x) \vert \pi_g) (f(x) \vert \pi_g) dx} = r^*} \int_{[0, 1]^N} d(x) (\mu(x)\vert \pi_g) \cdot\\
		& \ \ \ \ \ \ \ \ \ \ \ \ \ \ \ \ \ \ \ \ \ \ \ \ \ \ \ \ \ \ \ \ \ \ \ \ \ \ \ \ \ \ \ \ \left[ \left(f(x)\vert \pi_g\right)v_q - \left(1- f(x)\vert \pi_g\right) v_u\right] dx.
\end{align*}
The first-order condition then implies $d_g^*(x)$ is a cutoff function of
\begin{align*}
		& \frac{(\mu(x)\vert \pi_g) \left[ \left(f(x)\vert \pi_g \right)v_q - \left(1- f(x)\vert \pi_g\right) v_u\right]}{\frac{(\mu(x) \vert \pi_g) (f(x) \vert \pi_g)}{\int_{[0, 1]^N} (\mu(z) \vert \pi_g) (f(z) \vert \pi_g) dz}}\\
=\		&\pi_g \left[v_q - \frac{1 - \pi_g}{\pi_g \cdot \frac{p_q(x)}{p_u(x)}} \cdot v_u \right],
\end{align*}
which is comonotonic with the likelihood function. Thus, $d_g^*$ is a likelihood-cutoff decision policy. Equation \eqref{true_positive} implies different likelihood-cutoff decision policies generate different probabilities of acceptance, conditional on being qualified. So $d_A^*$ and $d_B^*$ must be the same likelihood-cutoff decision policy. This proves the second part of the lemma.

\end{proof}

%%%%%%%%%%%%%%%%%%%%%%%%%%%%%%%%%%%%%%%
%%%%%%%%%%%%%%%%%%%%%%%%%%%%%%%%%%%%%%%
%%%%%%%%%%%%%%%%%%%%%%%%%%%%%%%%%%%%%%%
%%%%%%%%%%%%%%%%%%%%%%%%%%%%%%%%%%%%%%%
%%%%%%%%%%%%%%%%%%%%%%%%%%%%%%%%%%%%%%%
%%%%%%%%%%%%%%%%%%%%%%%%%%%%%%%%%%%%%%%
%%%%%%%%%%%%%%%%%%%%%%%%%%%%%%%%%%%%%%%
%%%%%%%%%%%%%%%%%%%%%%%%%%%%%%%%%%%%%%%
%%%%%%%%%%%%%%%%%%%%%%%%%%%%%%%%%%%%%%%
%%%%%%%%%%%%%%%%%%%%%%%%%%%%%%%%%%%%%%%
%%%%%%%%%%%%%%%%%%%%%%%%%%%%%%%%%%%%%%%
%%%%%%%%%%%%%%%%%%%%%%%%%%%%%%%%%%%%%%%
%%%%%%%%%%%%%%%%%%%%%%%%%%%%%%%%%%%%%%%
%%%%%%%%%%%%%%%%%%%%%%%%%%%%%%%%%%%%%%%
%%%%%%%%%%%%%%%%%%%%%%%%%%%%%%%%%%%%%%%
%%%%%%%%%%%%%%%%%%%%%%%%%%%%%%%%%%%%%%%
%%%%%%%%%%%%%%%%%%%%%%%%%%%%%%%%%%%%%%%
%%%%%%%%%%%%%%%%%%%%%%%%%%%%%%%%%%%%%%%
%%%%%%%%%%%%%%%%%%%%%%%%%%%%%%%%%%%%%%%
%%%%%%%%%%%%%%%%%%%%%%%%%%%%%%%%%%%%%%%
%%%%%%%%%%%%%%%%%%%%%%%%%%%%%%%%%%%%%%%
%%%%%%%%%%%%%%%%%%%%%%%%%%%%%%%%%%%%%%%
%%%%%%%%%%%%%%%%%%%%%%%%%%%%%%%%%%%%%%%
%%%%%%%%%%%%%%%%%%%%%%%%%%%%%%%%%%%%%%%
%%%%%%%%%%%%%%%%%%%%%%%%%%%%%%%%%%%%%%%
%%%%%%%%%%%%%%%%%%%%%%%%%%%%%%%%%%%%%%%
%%%%%%%%%%%%%%%%%%%%%%%%%%%%%%%%%%%%%%%
%%%%%%%%%%%%%%%%%%%%%%%%%%%%%%%%%%%%%%%
%%%%%%%%%%%%%%%%%%%%%%%%%%%%%%%%%%%%%%%
%%%%%%%%%%%%%%%%%%%%%%%%%%%%%%%%%%%%%%%
%%%%%%%%%%%%%%%%%%%%%%%%%%%%%%%%%%%%%%%
%%%%%%%%%%%%%%%%%%%%%%%%%%%%%%%%%%%%%%%

\bibliographystyle{econ}
\bibliography{References}

\end{document}